\documentclass[aps,pre,twocolumn,10pt,amsmath,amsfonts,amssymb,floatfix]{revtex4-1}
\usepackage{amssymb}
\usepackage{amsmath,bm}
\usepackage{graphicx,color}
\usepackage{bbm}
\usepackage{multirow}
\newcommand{\B}[1]{{\bm{#1}}}

\begin{document}

\title{Mechanical Failure in Amorphous Solids: Scale Free Spinodal Criticality}

\author{Itamar Procaccia, Corrado Rainone and Murari Singh}

\affiliation{Department of Chemical Physics, the Weizmann Institute of Science, Rehovot 76100, Israel}

\begin{abstract}
The mechanical failure of amorphous media is a ubiquitous phenomenon from material engineering to
geology. It has been noticed for a long time that the phenomenon is ``scale-free", indicating some
type of criticality. In spite of attempts to invoke ``Self-Organized Criticality", the physical
origin of this criticality, and also its universal nature, being quite insensitive to the nature
of microscopic interactions, remained elusive. Recently we proposed that the precise nature of
 this critical behavior is manifested by a spinodal point of a thermodynamic phase transition. Moreover, at the spinodal point there exists a divergent
correlation length which is associated with the system-spanning instabilities (known also as shear
bands) which are typical to the mechanical yield. Demonstrating this requires the introduction of an `order parameter' that is suitable for distinguishing between
disordered amorphous systems, and an associated correlation function, suitable for picking up the growing correlation length.
The theory, the order parameter, and the correlation functions used are universal in nature and can be applied
to any amorphous solid that undergoes mechanical yield. Critical exponents for the correlation length divergence and the system
size dependence are estimated. The phenomenon is seen at its sharpest
in athermal systems, as is explained below; in this paper we extend the discussion also to thermal
systems, showing that at sufficiently high temperatures the spinodal phenomenon is destroyed by thermal
fluctuations.
\end{abstract}

\maketitle

\section{Introduction}

Mechanical failure of amorphous solids is an unwanted and often catastrophic event, occurring when enough strain and stress accumulate due to external loading. The phenomenon is ubiquitous in nature in the form of earthquakes due to tectonic activity and in material engineering due to shear or tensile strains. The phenomenon is
known to be ``scale-free" in the sense that the statistics of energy release upon failure appears to have
no typical scale, a characteristic that is exemplified by the Gutenberg-Richter law~\cite{Gutenberg13} in the geophysical context. Many authors commented that this scale-free nature indicates that material failure should be a critical phenomenon with power-law scaling, but until recently the precise origin and the actual character of this criticality remained unknown. Precisely three decades ago P. Bak and coworkers~\cite{BTW87} offered the idea of ``Self Organized Criticality" to
explain the ubiquity of such scale-free statistics, but the correspondence to the microscopic structure
of amorphous solids and the particle-scale mechanisms that are responsible for the phenomenon remained
mysterious. Recently \cite{16JPRS,17PPRS} the source of the criticality was revealed in the form of a {spinodal criticality} which appears to be quite universal in athermal conditions independently {from} the detailed microscopic interactions between the particles forming the amorphous solid. This criticality is not at all `self-organized', rather it is forced on the system by the external loading. The aim of this paper is to review the pertinent
features of this phenomenon and extend its exploration from athermal systems to amorphous solids at finite temperatures. Among other issues discussed below it will be shown that when the temperature becomes high enough
the spinodal characteristics are destroyed by thermal fluctuations.

Solids are states of matter capable to respond elastically to a small externally applied shear deformation~\cite{59LL}. However when the external strain grows the response of all solids becomes
 mixed with plastic deformations, and eventually they suffer a mechanical yield. In crystalline
  solids plasticity and yield involve defects and dislocations. In amorphous materials such as molecular and colloidal glasses, foams, and granular matter there is no long range order with respect to which defects {can be}
  defined. Thus the mechanisms of plasticity and yield {in amorphous materials} need {be} understood along different lines from those of
  crystalline matter. The physics near the yielding point of this vast class of materials, as reported in a host of strain-controlled simulations~\cite{04VBB,04ML,05DA,06TLB,06LM,09LP,11RTV} and experiments~\cite{06SLG,13KTG,13NSSMM} shows a high degree of universality despite the different nature of the systems involved. Importantly, one finds that at the onset of flow at yielding, there appear
  typical system spanning excitations referred to as shear-bands~\cite{11BB,12DHP}. We refer to a plastic event as a shear band when
   previously homogeneous shear strongly localizes, leaving the rest of the material less perturbed. This phenomenon is of capital importance for engineering applications as it is responsible for the brittleness typical of glassy materials, in particular metallic glasses~\cite{06AG}, whose potential for practical use is stymied by their tendency to shear-band and fracture~\cite{12DHP,13DHP,13DGMPS}. Measurements of
   plastic events occurring after yield reveal scale-free energy or stress drops, typically characterized
   by power-law statistics~\cite{KLP10b,FS17}. The aim of this paper is to present the current understanding of this scale-free behavior which, as said above, is suspected to be related to some criticality.

  This paper is organized as follows: in Sect. \ref{yield} we discuss the universal features of mechanical yield, explaining that
   any appropriate theory must use generic order parameters which are equally applicable to a large
   variety of amorphous solids. This is crucial. After introducing the ``overlap" order parameter, we turn
   to using it to investigate the physics of yield in athermal
   conditions. The key result will be that yield is tantamount to a spinodal point in the emerging
   phase transition that is associated with the phenomenon. In Sect. \ref{criticality} we follow up on
   the identification of the precise criticality that is implied by the spinodal point, and we study
   the correlation functions that are expected to exhibit a divergent correlation length. We {then show in Sect \ref{numerics}}
   that the correlation length {associated to these correlators} diverges as a power law in the distance from the spinodal point, cf. Eq.~(\ref{xilaw}) below. Section \ref{T}
   explores the modifications caused by having a finite temperature. Not too surprisingly, we will discover
   that at {sufficiently} higher temperatures, fluctuations destroy the spinodal characteristics, forcing a cross
   over to different statistics of the energy drops. In Sect. \ref{summary} we offer a summary
   of the paper and thoughts about the road ahead. We also comment on the notion of `self organized criticality' which
   is a very vague notion and explain how it is related to the results of this paper.

\section{Mechanical Yield, Universality and Order Parameter}
\label{yield}
\begin{figure}
\includegraphics[width=0.5\textwidth]{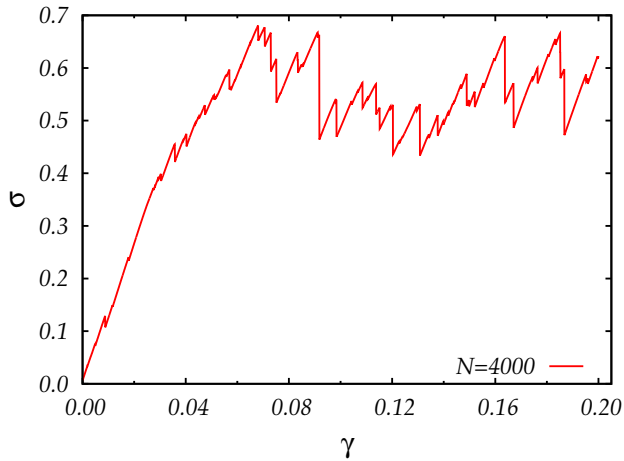}
\caption{A typical stress vs. strain curve resulting from a shear loading of an
amorphous solids using an AQS protocol. Similar transitions between a regime in which
the stress rises on the average as a function of strain to a second regime after
a yield point $\gamma_{_{\rm Y}}$ have been observed in a countless experiments and
simulations, requiring an explanation using a generic theory that is insensitive to
microscopic details.}
\label{svss}
\end{figure}

\subsection{Universality of Mechanical Yield}
To introduce the main issue consider Fig.~\ref{svss} showing a typical stress vs. strain curve obtained using standard numerical simulations in a strain-controlled athermal quasistatic (AQS)
shearing protocol. This particular figure pertains to a Kob-Andersen 65-35\% Lennard Jones Binary Mixture~\cite{94KA} of 4000 particles in $2d$. Similar curves were computed and measured in a large variety of simulations and experiments. The universal features that need to be observed are {the following}: (i) For very small strain values the stress increases linearly
according to the laws of linear elasticity. One should note that the region of purely elastic
behavior is expected to reduce with the system size, shrinking to nonexistence in the thermodynamic
limit \cite{11HKLP}. Nevertheless, before a value of the strain known as the ``yield strain" $\gamma_{_{\rm Y}}$, the plastic
events are ``small", in the precise sense that the energy drop $\Delta U$ associated with them is system
size independent
\begin{equation}
\Delta U \sim N^0 \ ,
\label{UN0}
\end{equation}
where $N$ is the number of particles in the systems. The nature of these plastic events is identified
as quadrupolar displacements, known also as Eshelby~\cite{E57} events, which can release stress locally
in regions that are particularly susceptible to the type of loading employed. The important
{point} is that, whether elastically or punctuated by plastic events, the stress $\sigma$ continues to increase
with the strain $\gamma$ until the latter exceeds $\gamma_{_{\rm Y}}$ which in Fig.~\ref{svss} is about
$\gamma_{_{\rm Y}}\approx 0.07$. After that point, in strain
controlled protocols the strain increases without increasing the average stress - the material ``flows"
keeping an average ``flow stress". In stress controlled experiments, exceeding the average flow stress results
in a mechanical collapse of the material. In athermal conditions it was found that the transition
around $\gamma=\gamma_{_{\rm Y}}$ is associated with a change in the plastic response which is no
longer localized, but rather exhibits system spanning events, known also as micro shear bands,
in which the energy release becomes sub-extensive~\cite{KLP10b},
\begin{equation}
\Delta U \sim N^\beta \ , \quad \beta=2/3 \ .
\label{UN}
\end{equation}
The mechanism for the creation of these micro shear bands was elucidated in~\cite{12DHP}, {and} it has to do with the
preferred appearance of concatenated series of Eshelby quadrupoles (lines in $2d$ or embedded in a plane
in $3d$) that organize the displacement field to localize the shear on narrow lines or planes respectively.
The interested reader is referred to Ref.~\cite{13DHP} where detailed energy estimates were offered to explain
the energetic preference of single Eshelby quadrupoles at low strains vs. the appearance
of a density of such objects at higher values of the strain.

The key observation is that after yield strain $\gamma_{_{\rm Y}}$ the stress cannot grow on the average, no matter how much the strain is increased. What remained obscure for a long time is what is the difference in the material before and after the yield point; why the stress could continue growing with the strain before yield, but it cannot do that after yield. Since the phenomenon is ubiquitous, the universality of this basic phenomenology of yielding begs an explanation in terms of a universal theory, in the sense that such a theory should rely on a statistical-mechanical framework and be independent of details such as chemical composition and the production process of the material.

\subsection{Order Parameter and Transition}
\label{transition}

In Ref.~\cite{16JPRS} it was made clear that the difficulty in making a distinction between the
pre- and post-yield configurations lies in the fact that {\em there is really no distinction}.
The crux of the matter is not in the nature of configurations but in their number. The yield
takes place because of a sudden opening up of a vast number of marginally stable configurations that are not
at the system's disposal before yield. To demonstrate this one needs to employ an order
parameter that is designed \cite{FP95,10CCGGGPV,12YM,13Ber,14BC,14BCTT,15BJ,15NBC} to compare two different glassy configurations $\{\B r^{(1)}_i\}_{i=1}^N$ and $\{\B
r^{(2)}_i\}_{i=1}^N$
\begin{equation}
Q_{12} \equiv \frac{1}{N}\sum_i^N\theta(a-|\B r^{(1)}_i - \B r^{(2)}_i|)\ ,
\label{Q12}
\end{equation}
wherein $\theta(x)$ is the Heaviside step function. The parameter
$a$ is of the order of the microscopic interaction length, and is determined by trial and error. The quantity $Q_{12}$  is
called an ``overlap" since it has a value that goes from $0$ (completely
decorrelated configurations) to $1$ (identical particle coordinates within the tolerance of $a$). Its purpose is to measure the degree of similarity between configurations.

Let us now consider a glass, made by quenching a super-cooled liquid with $N$ particles down to a
certain temperature $T\ge 0$ at a suitable rate. A glass is an amorphous solid
wherein particles vibrate around an amorphous structure. So, if we take two
configurations $\{\B r_i^{(1)}\}_{i=1}^N$ and $\{\B r_i^{(2)}\}_{i=1}^N$ from this glass at two
different times, they will be most
likely close to each other with $Q_{12}$ of the order of unity. If one is able
to obtain a good sampling of the typical configurations visited by the particles
in the glass, one can measure the probability distribution of the overlap
$P(Q_{12})$, which will be strongly peaked around an average value $\langle
Q_{12}\rangle$ close to unity. The configurations visited by the particles will
then form a small connected ``patch" in the configuration space of the system,
selected by the amorphous structure provided by the last configuration that was
realized by the liquid glass former before it fell out of equilibrium while forming
a glass.

Things will change once we begin straining this glass. While the stress increases,
there appear plastic events that are associated with irreversible displacements in
the particle positions. The average order
parameter $\langle Q_{12}\rangle(\gamma)$ responds
to these displacements, reducing from $O(1)$ to lower values. An important
point to understand is that before reaching the yield strain
$\langle Q_{12}\rangle(\gamma)$ tends to remain around unity, but as the
mechanical yield takes place a sharp phase transition occurs, whereupon sub-extensive
plastic events \cite{09LP,12DHP,13DGMPS} begin to take place. These
are sufficiently large, cf. Eq.~(\ref{UN}), to cause
substantial displacements, allowing different
regions of the configuration space to affect the order parameter.
In such a situation, the distribution $P_\gamma(Q_{12})$ may develop two peaks: one at
high $Q_{12}$ corresponding to configurations in the same patch and one
for a smaller value of $Q_{12}$ corresponding to configurations that were ``ergodized"
by the mixing of the sub-extensive plastic events.

To demonstrate this fundamental idea we can use any model glass, since this
order parameter description is expected to be universal. Here we review
molecular dynamics simulations
of a Kob-Andersen 65-35\% Lennard Jones (LJ) Binary Mixture in $2d$, using five system
sizes, $N=500, 1000, 2000, 4000$ and $N=10000$. We chose $Q_{12}$ with $a = 0.3$ in LJ
units, but verified that changes in $a$ leave the emerging picture invariant.
As a first step, we prepared a glass by equilibrating the system at $T=0.4$, and
then quenching it (the rate is $10^{-6}$ in LJ units) down to $T=1\cdot10^{-6}$ into a
glassy configuration. The sample is then heated up again to $T=0.2$, and a starting
configuration of particle positions is chosen at this temperature. Note that while at
$T=0.4$ equilibration is sufficiently fast, at $T=0.2$ the computation time is much shorter
than the relaxation time. The
configuration is then assigned a set of velocities randomly drawn from the
Maxwell distribution at $T=0.2$, and these different samples are then quenched down to
$T=0$ at a rate of $0.1$. This procedure can be repeated any number of times (say $n$ times),
and it allows us to get a sampling of the configurations inside one single ``patch". We
verify that the typical overlap of the ensemble of configurations so
obtained in one patch is close to $\langle Q_{12}\rangle = 1$, signaling that indeed the ensemble
is completely located in a single patch.

Having generated one such patch, we repeat the
procedure starting from another equilibrated configuration of the liquid to create another patch.
The process is then repeated to generate as many patches (say $m$ patches) as needed to obtain good
statistics, depending on the system size.
\begin{figure}
\includegraphics[scale = 0.50]{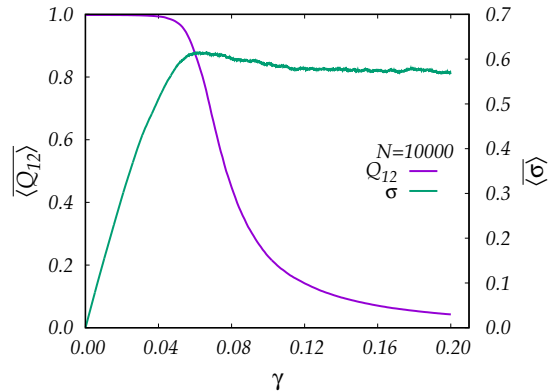}
\caption{The averaged order parameter $\overline{\langle Q_{12}\rangle}$ as a function
of $\gamma$ (left scale) and the averaged stress as a function of $\gamma$ (right scale).
The averaging is over all the patches for this system size $N=10000$. Note the phase transition
that occurs near the yield strain $\gamma_{_{\rm Y}}$. The transition gets sharper with the system size, see
Fig.~\ref{plots} and the associated discussion below.}
\label{Qvsgamma}
\end{figure}

We then apply to {\em each configuration} in a given patch an AQS protocol as described above.
This will create for each value of $\gamma$ a \emph{strained ensemble} of
configurations in the patch. The order parameter Eq.~(\ref{Q12}) is
computed by using {\em all} the $n(n-1)/2$ unique pairs of configurations generated in the strained
ensemble at a given $\gamma$. We stress that we do {\em not} compare configurations at a given
value of $\gamma$ to the reference configuration at $\gamma=0$, but rather the overlap between pairs
of configurations at the same value of $\gamma$. Having computed the $\gamma$ dependence of an average $\langle Q_{12}\rangle$ from
the $n(n-1)/2$ of configurations in one patch, we average the results over $m$ patches to obtain
the average order parameter denoted as $\overline{\langle Q_{12}\rangle}$, wherein the acute brackets denote the average
over a single patch and the overline denotes the average over all patches. We present the
results for $N=10000$ in Fig.~\ref{Qvsgamma}. Note that the initial
ensemble for $\gamma = 0$ shows a value of the averaged order parameter $\overline{\langle Q_{12}\rangle} = 1$, signifying
that our initial ensemble is indeed composed of close-by configurations. As the ensemble is
strained, the value of the order parameter gets lower, dropping towards zero when the strain
is increased beyond the yield strain. Below we will show that the sharpness of the transition depends
on the system size $N$, getting sharper and sharper when $N$ increases, {as expected}.

\begin{figure}
\vskip -0.5 cm
\includegraphics[scale = 0.65]{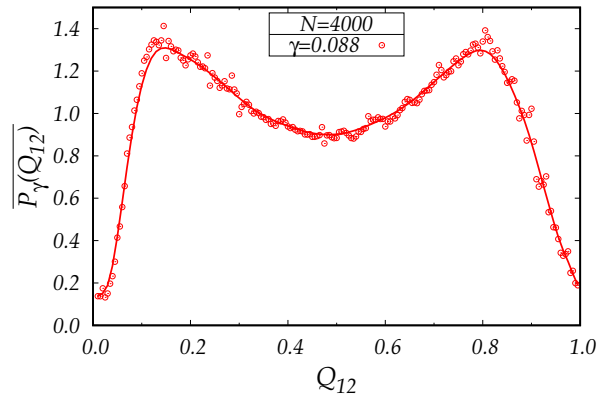}
\caption{The probability distribution function $\overline{P_\gamma(Q_{12})}$ at $\gamma_{_{Y}}=0.088$ averaged over
100 initial configurations each of which has 500 different realizations to obtain $\overline{P_\gamma(Q_{12})}$. At this value of the strain the pdf has two peaks of equal height. We identify this value of $\gamma$
as the point of the phase transition.}
\label{yield-point}
\end{figure}

To determine the yield strain $\gamma_{_Y}$ accurately, one should construct the probability distribution function (pdf) $P_\gamma(Q_{12})$ by hystogramming the values of $Q_{12}$ within a patch of $n$ configurations obtained as explained above, and then average the result over the $m$ available patches. The result is denoted
$\overline{P_\gamma(Q_{12})}$.
We ask at which value of $\gamma$ this averaged pdf has two equally high peaks, see Fig.~\ref{yield-point}. The resulting $\overline{ P_\gamma(Q_{12})}$ determines a value of $\gamma_{_{Y}}\simeq 0.088$. Note that this criterion implies a sharp definition of ``yield" which seems absent in the current literature. If accepted, it indicates that the mechanical yield occurs beyond the stress overshoot in correspondence with the mean-field results of Ref.~\cite{15RUYZ}. We should also state here
the yield point and the spinodal point (denoted below as $\gamma_S$) are not identical for finite $N$, although they become
closer when $N$ increases, and see below for details.

\begin{figure}
\includegraphics[scale = 0.65]{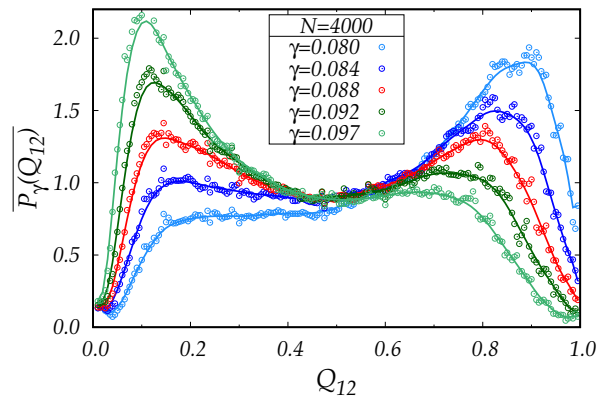}
\caption{The probability distribution function $\overline{P_\gamma(Q_{12})}$ in the vicinity
of the critical point $\gamma_{_{Y}}=0.088$}
\label{trans}
\end{figure}

Once we identify the phase transition point, we can demonstrate the transition itself.
In Fig.~\ref{trans} we display the change in $\overline{P_\gamma(Q_{12})}$ in the vicinity of the critical
point $\gamma_{_{Y}}$ as a function of $\gamma$.
Within a very narrow range of $\gamma$, of the order of $\Delta \gamma \approx 0.017$, we observe a first-order
like transition from
a pdf with dominant peak at high values of $Q_{12}$ to a dominant peak at low values of $Q_{12}$.   We capture a very
unambiguous and qualitative change in behavior as the yielding point is reached.

\begin{figure}
\includegraphics[scale = 0.65]{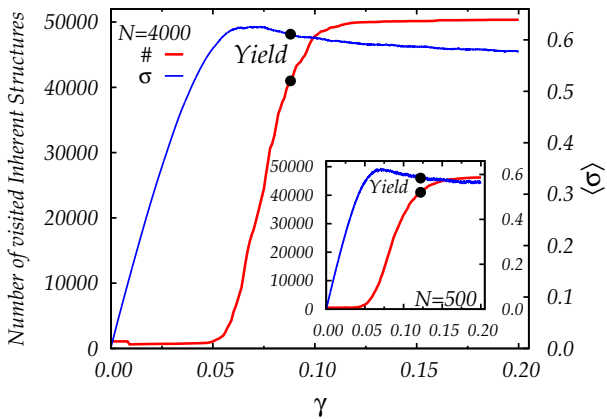}
\caption{The number of configurations which pass below the threshold value $Q_{12}=0.8$ of the overlap order
parameter as a function of
the strain $\gamma$ for $N=4000$. In the inset we show the same test for $N=500$. The conclusion is that {\em all the configurations} lose the mutual overlap in the vicinity of the yield point $\gamma_{_Y}$}.
\label{number}
\end{figure}

To sharpen the understanding of what is happening in the vicinity of the yield point we examine next {\em how many} of our realizations loose the tight overlap  and where the loss of
overlap is taking place. To this aim we consider, as an example for the system of 4000 particles, all the 50,000 realizations that we have from 100 patches each containing 500 configurations. These are obtained by 100 choices of liquid realizations, each of which is velocity randomized 500
times (chosen with Boltzmann probabilities).  When the strain $\gamma$ is increased in our AQS algorithm, we keep
computing the order parameter $Q_{12}$ where the first configuration $\{\B r^{(1)}_i\}_{i=1}^N$ in Eq.~(\ref{Q12}) is chosen randomly from all the available configurations at that value of $\gamma$,
and the second is any one of the other available configurations at the same value of $\gamma$. We confirmed that
changing the randomly chosen $\{\B r^{(1)}_i\}_{i=1}^N$ does not affect the results. Next, choosing $Q_{12}=0.8$ as a threshold value, we now count how many of our observed configurations cross this threshold and exhibit $Q_{12}\le 0.8$. The number of configurations that do so as a function of the strain (superimposed on the stress vs. strain curve) is shown in Fig~\ref{number}.
The conclusion of this test is that in the vicinity of the yield point $\gamma_{_Y}$ {\em all the configurations}
lose their overlap with the initial configuration, but {\em not before}. The mechanical yield is tantamount to the opening up of a vast number of possible configurations, whereas before yield the system is still constrained to
reside in the initial meta-basin of the free energy landscape.

The upshot of these results is that we are able to focus on the essential feature that
is responsible for the mechanical yield: a
very constrained set of configurations
available to the system before yield is replaced upon yield with a vastly larger
set of available configurations. This much larger set is generic; we would like to refer to the
phenomenon as ``stressed ergodization". The initially prepared close-by configurations are now scattered,
but all of them are stressed with stress value close to the yield stress. They are all marginally
stable in the sense that they would yield plastically with any
increase of strain \cite{10KLP,15HJPS}.  We propose this as a universal
mechanism for the ubiquitous prevalence of stress vs. strain curves that look so similar in a huge variety
of glassy systems.

\subsection{System size dependence}

\begin{figure}
\includegraphics[scale = 0.55]{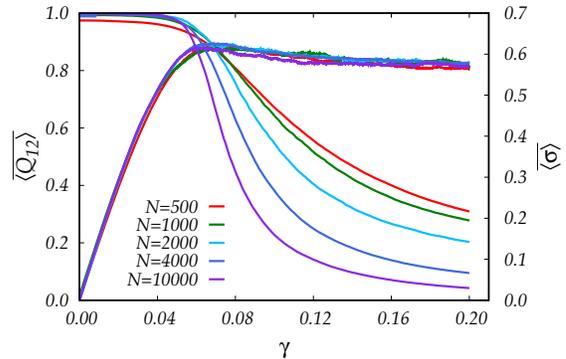}
\caption{Demonstrating the sharpening of the transition with the system size.}
\label{plots}
\end{figure}

\begin{figure}
\includegraphics[scale = 0.60]{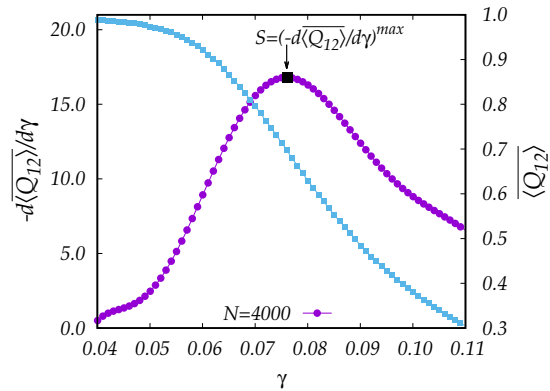}
\includegraphics[scale = 0.60]{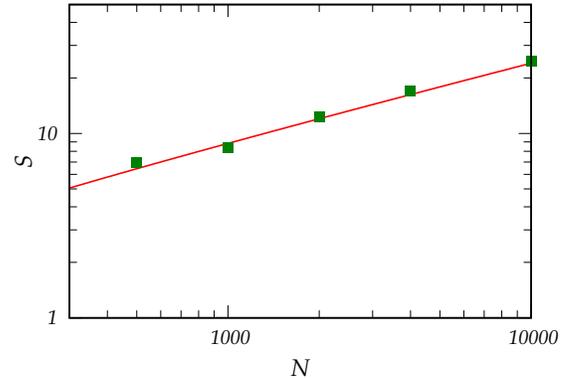}
\caption{Upper panel: A typical graph of $\overline{\langle Q_{12}\rangle}$ as a function of $\gamma$
(here for $N=4000$ (right scale) and the slope of the same function (left scale). Lower panel: The maximal slope of the function $\overline{\langle Q_{12}\rangle(\gamma)}$ as a function of system size.}
\label{sharpness}
\end{figure}

In the context of first-order phase transitions one expects that the transition should become
sharper as a function of system size. To this aim we consider the dependence of $\overline{\langle Q_{12}\rangle}$ on
$\gamma$ for a series of system sizes, see Fig.~\ref{plots}.
Indeed, the sharpening of the transition
is obvious to the bare eye. To quantify it we evaluated the derivative of this function, see the upper panel
in Fig.~\ref{sharpness} for $N=4000$, and computed the maximum of this derivative function, denoted
as
\begin{equation}
S\equiv \underset{\gamma}{\rm max} \left(-\frac{d\overline{\langle Q_{12}\rangle}}{d\gamma}\right)\ .
\label{SN}
\end{equation}
 Finally, the value of $S(N)$ is plotted in a log-log plot vs. the system size $N$ as shown in the lower panel
of Fig.~\ref{sharpness}.
This log-log plot indicates the existence of a power law of the form
\begin{equation}
S\approx C N^\theta \ ; \theta=0.41\pm 0.09 \ .
\label{SvsN}
\end{equation}
The error bars measured here suggest that the exact value of the exponent $\theta$ is $\theta=1/2$.
Such an exponent indicates that the width of the transition is not determined by the thermal fluctuations
in the parent fluid from which our glassy patches were quenched, {as it would be in the case of an ordinary first-order transition}. Rather, it is dominated by the
disorder fluctuations ({i.e. sample to sample fluctuations, due to the fact that each glass is randomly ``selected`` at quenching time by a parent configuration in the high temperature liquid. This causes $\gamma_{_{\rm Y}}$ to vary from sample to sample. To test this
hypothesis we return to our numerical data and compute, for each patch, a yield point $\gamma_c$
which we identify as the first value of the strain for which $\langle Q_{12}\rangle \le 0.5$. Having done so
we can evaluate the probability distribution function $P(\gamma_c,N)$. These functions obviously depend on the systems size as shown in the upper panel of Fig.~\ref{Pyield}.
\begin{figure}
\includegraphics[scale = 0.60]{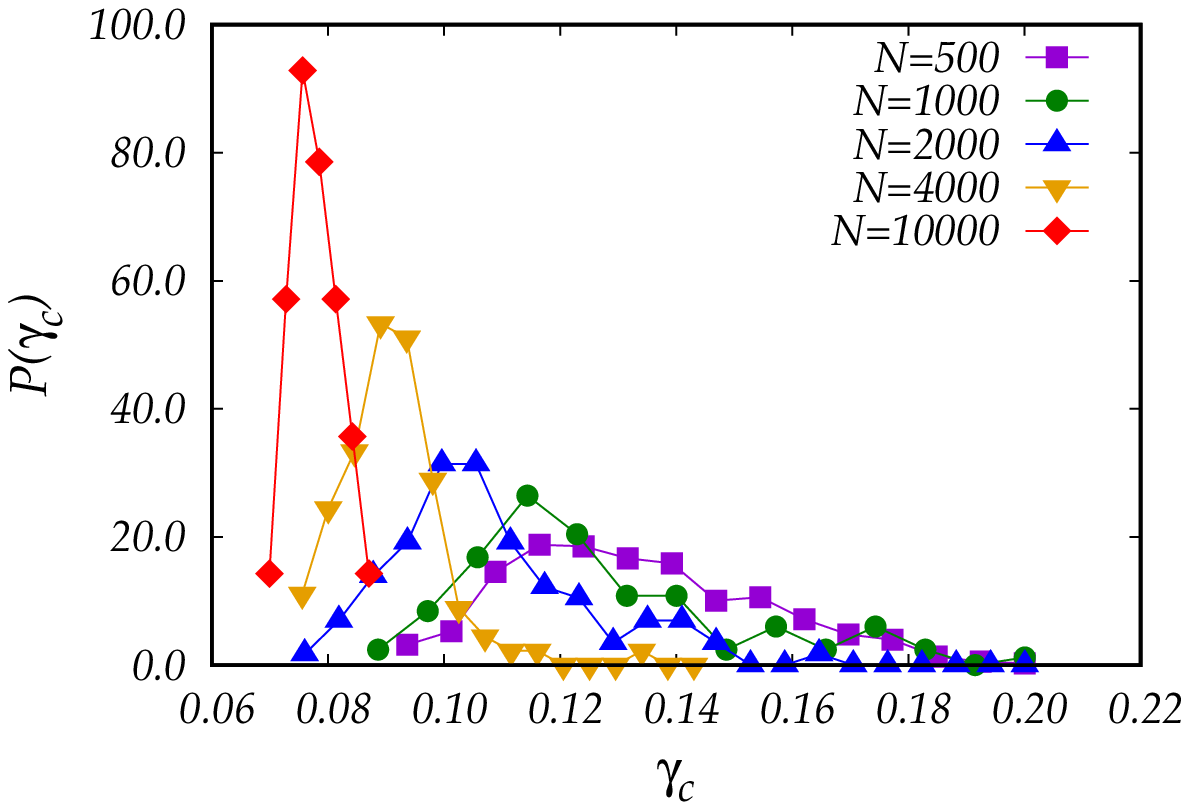}
\includegraphics[scale = 0.60]{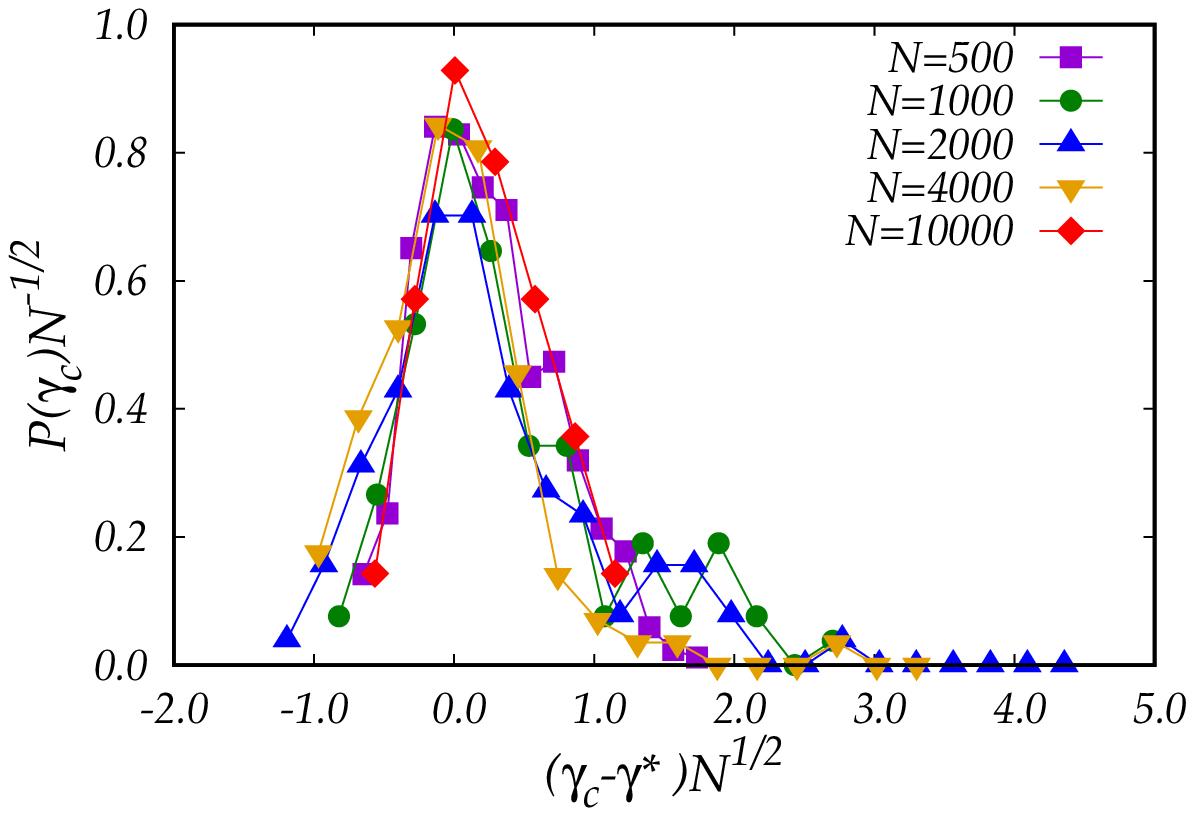}
\caption{Upper panel: the pdf's $P(\gamma_c,N)$ for different systems sizes from $N=500$
to $N=10000$. Lower panel: data collapse upon rescaling the pdf's $P(\gamma_c,N)$ according
to Eq.~(\ref{rescalepdf}) to obtain a scaling function $\tilde P(x)$ with $x=\left((\gamma_c-\gamma^*)\sqrt{N}\right)$.}
\label{Pyield}
\end{figure}
To examine the scaling of the width of these distributions we rescale the data according to the
ansatz
\begin{equation}
P(\gamma_c,N) = \sqrt{N} \tilde P\left((\gamma_c-\gamma^*)\sqrt{N}\right)
\label{rescalepdf}
\end{equation}
where $\gamma^*$ is the peak value of each pdf.
The data collapse means that indeed the disorder
leads to a spread $\Delta_{\gamma_c}$ in the values of $\gamma_{_{\rm Y}}$ that scales like
\begin{equation}
\Delta_{\gamma_c} \sim N^{-1/2} \ ,
\end{equation}
which will end up as the scaling law Eq.~(\ref{SvsN}) with $\theta=1/2$. If we just had a thermal origin to the measured width
we could expect rather a scaling law with $\theta=1$, as typical of first-order transitions~\cite{BL84}. {This finding highlights the
pivotal role played by the fluctuations over the disorder in the finite-size scaling of the yielding transition.}
\subsection{Concluding this section}

The upshot of this section is that the yield is associated with a first order phase transition
such that before yielding the amorphous system is limited to a small patch in the configuration space, very far from any kind of ergodicity. The yielding transition is an opening of a much larger available configuration space, whereupon the system is ergodized subject to the constraint of constant mean stress. The generic configurations
that are created by the mixing caused by micro shear-bands include many marginally stable states which
yield easily upon the increase of strain. This is why the stress cannot increase further on the average.

This realization does not explain yet where is the criticality. In general first order phase transitions
are not characterized by diverging correlation lengths, while critical points associated with
second order phase transitions do. The point to understand, as sharpened in the next section, is
that first order phase transitions are bordered by spinodal points which do exhibit criticality. To see this
pictorially examine again Fig.~\ref{trans} and focus on the pdf associated with $\gamma=0.097$. At that
point the maximum of high values of $\overline{\langle Q_{12}\rangle}$ has been reduced to a saddle. This is a
spinodal point that we denote as $\gamma_S$ where the slope of the curve vanishes as well as the second derivative. This is
where a correlation length is expected to diverge as we are going to explain in the next section.
The reader should also take into account that when $N\to \infty$ also $\gamma_{_Y}\to\gamma_c\to \gamma_S$.
\begin{figure}
 \includegraphics[width=0.40\textwidth]{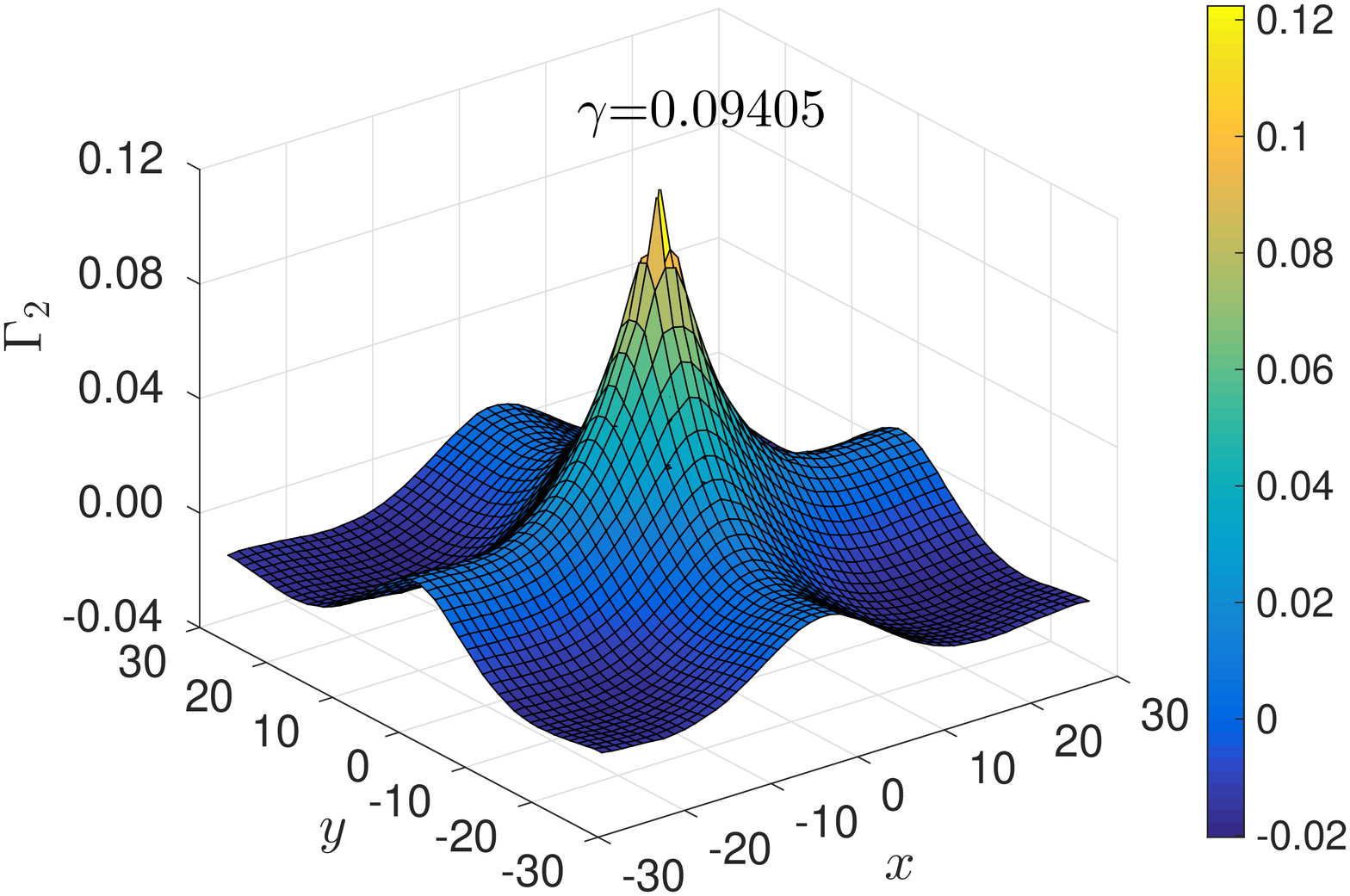}
 \includegraphics[width=0.40\textwidth]{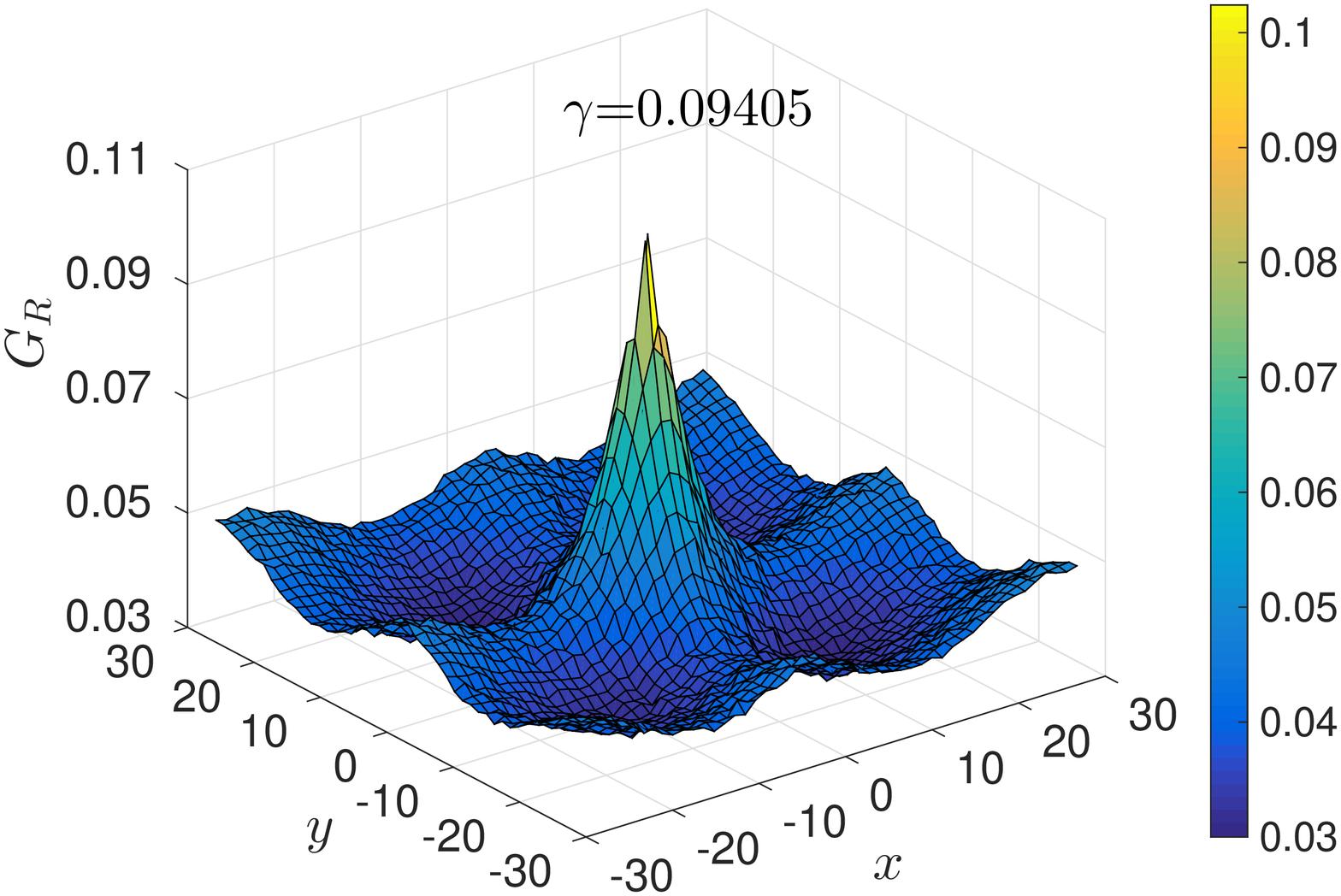}
 \includegraphics[width=0.40\textwidth]{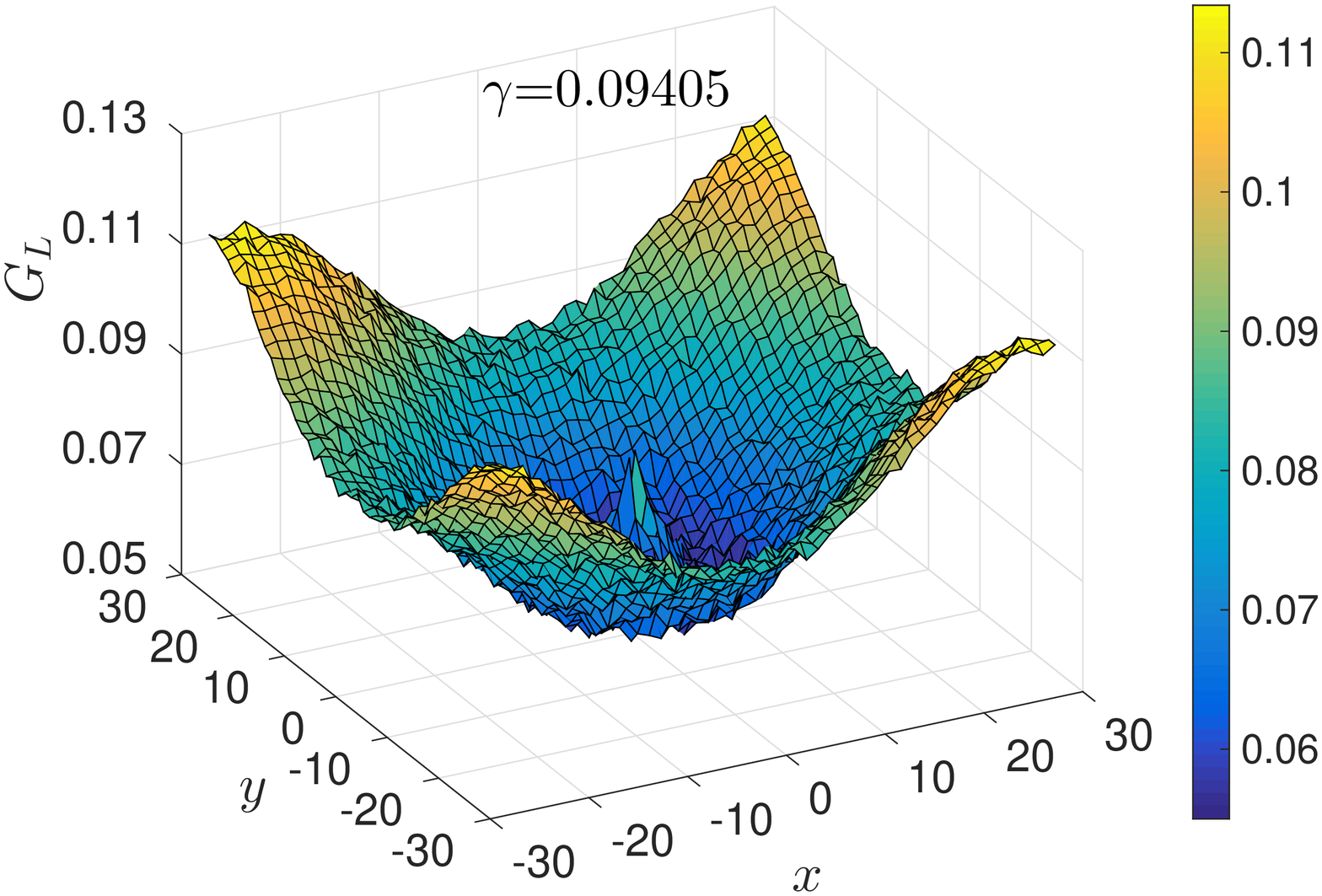}
 \caption{A 3-dimensional projection of the three correlation functions as a function of $x,y$.}
 \label{correlations}
 \end{figure}
\section{Theory of spinodal criticality}
\label{criticality}
The aim of this section is to clarify the identification of the yielding transition  as a spinodal point~\cite{16UZ}. This is the point where the metastable, high overlapped glassy patch of configurations, becomes unstable with respect to a new phase with low $Q_{12}$, associated with a stressed ergodized system in the presence of disorder~\cite{16NBT}. A previously known example of such a spinodal is the Mode Coupling crossover~\cite{11BB}, characterized by dynamical slowing down and heterogeneities, whose behavior is characterized by a \emph{dynamical lengthscale} which can be extracted from suitable multi-point correlators~\cite{11BB}. This kind of critical behavior should also be found at the yielding transition, conditional that one is able to derive the expression of the right correlator to measure. It is important to stress here that the reason that a spinodal point can be exposed and measured is that the
glassy time scales and the athermal conditions stabilize the metastable system until the spinodal point is crossed and the system becomes unstable against constrained ergodization. We will see below how thermal fluctuations
may destroy the spinodal characteristics.

In statistical mechanics with a suitable Gibbs free energy $G[\phi]$, {$\phi$ being the order parameter of choice,} stable phases are identified with its points of minimum in $\phi$. Of particular interest are instances for which the curvature of these minima goes to zero, inducing a critical behavior which manifests diverging susceptibilities-fluctuations, critical slowing down of the dynamics, and growing correlation lengths~\cite{02Z-J}. At a spinodal point, for example, one such minimum becomes unstable and transforms into a saddle. In the case of the order parameter $Q_{12}$ the general form of the free energy $s[Q_{12}]$ had been already derived and studied (see~\cite{98DKT} for a review) in the context of the theory of replicas originally developed for the study of spin-glasses, and its properties, at least at mean-field level, are well known (we refer to~\cite{15RUYZ,16RU} for the derivation of $s[Q_{12}]$ in the specific case of mean-field hard spheres); the matrix of second derivatives (or, using a more field-theoretic terminology, the mass matrix) is not diagonal in the basis of $Q_{12}$, and after diagonalization is found to have only three distinct modes, or masses~\cite{98DKT}. Of these, the most relevant ones are the so called \emph{replicon mode} $\lambda_R$, which for example goes to zero at the newly proposed Gardner transition~\cite{14CKPUZ}, and the \emph{longitudinal mode} $\lambda_L$ which is instead related to spinodal points \cite{16RU,16UZ} such as our yielding transition.  In Appendix \ref{theory} of this paper we review briefly the background theory
that is at the basis of the present approach.
\begin{figure}
 \includegraphics[width=0.4\textwidth]{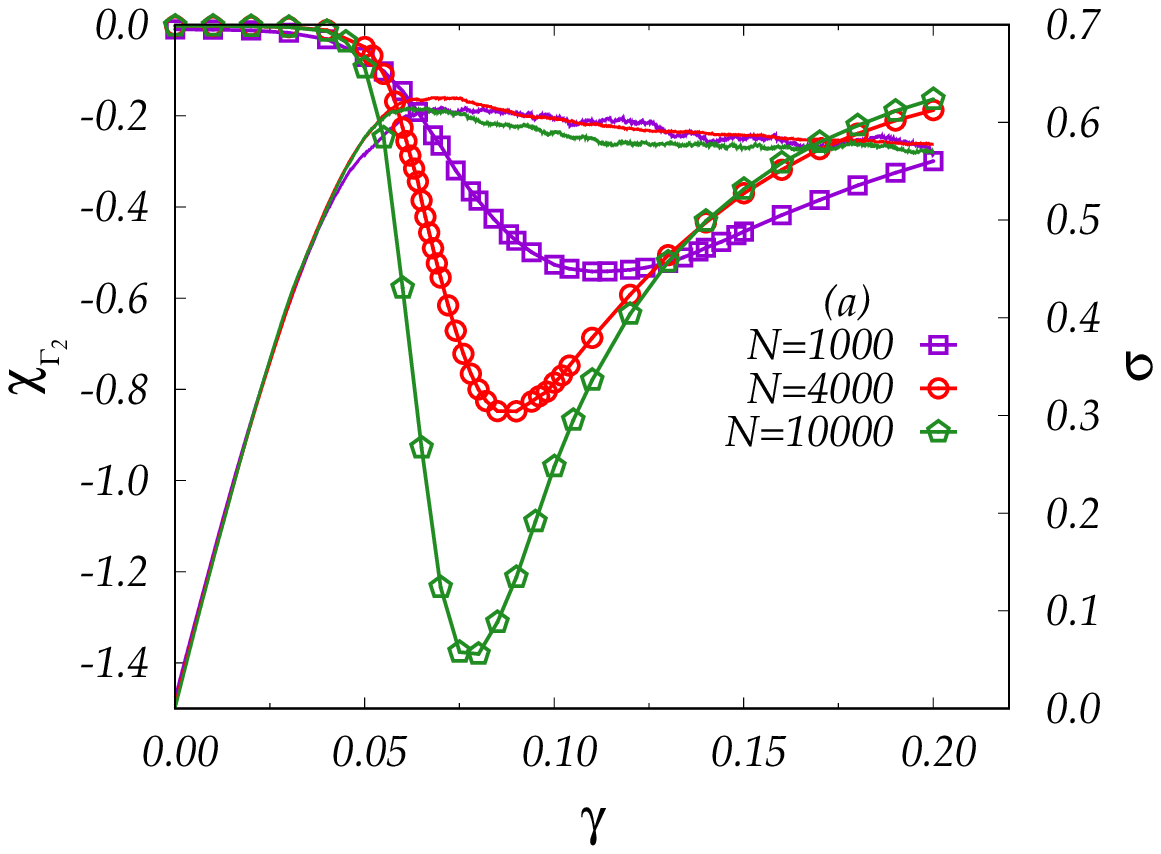}
 \includegraphics[width=0.4\textwidth]{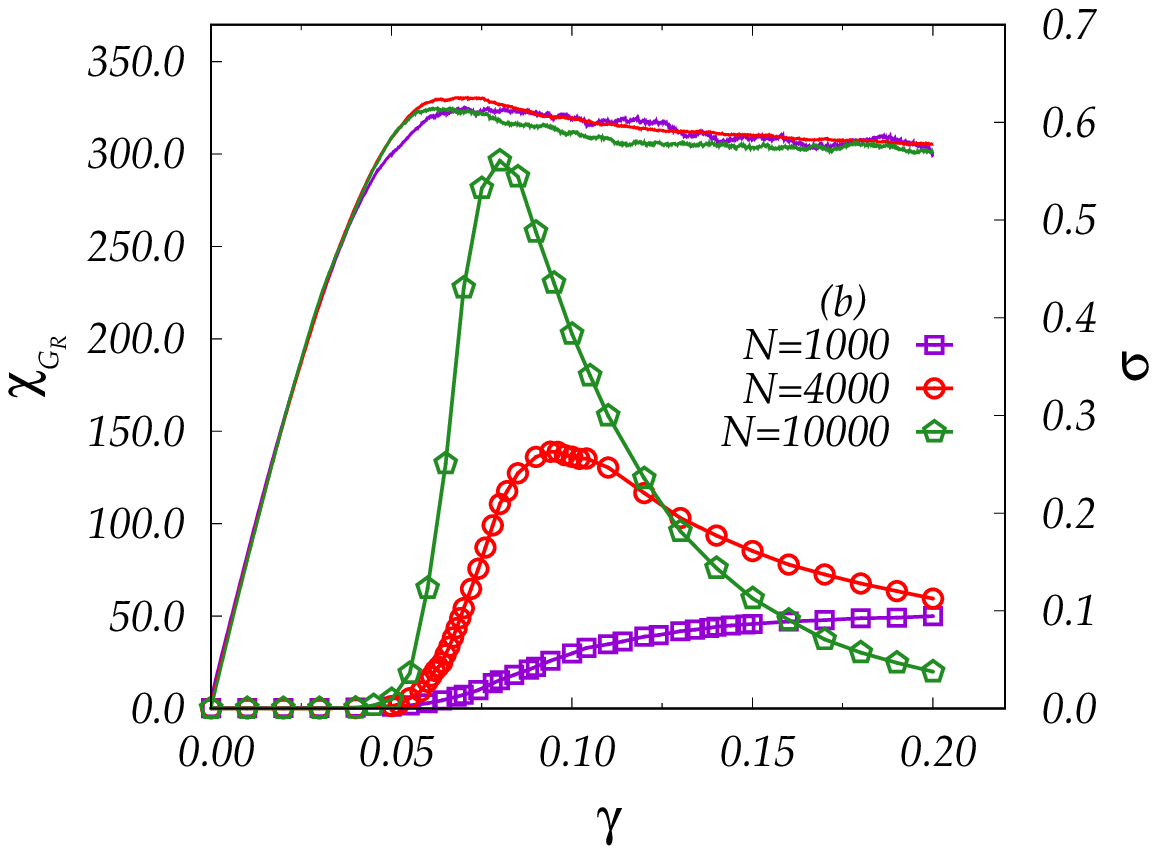}
  \includegraphics[width=0.4\textwidth]{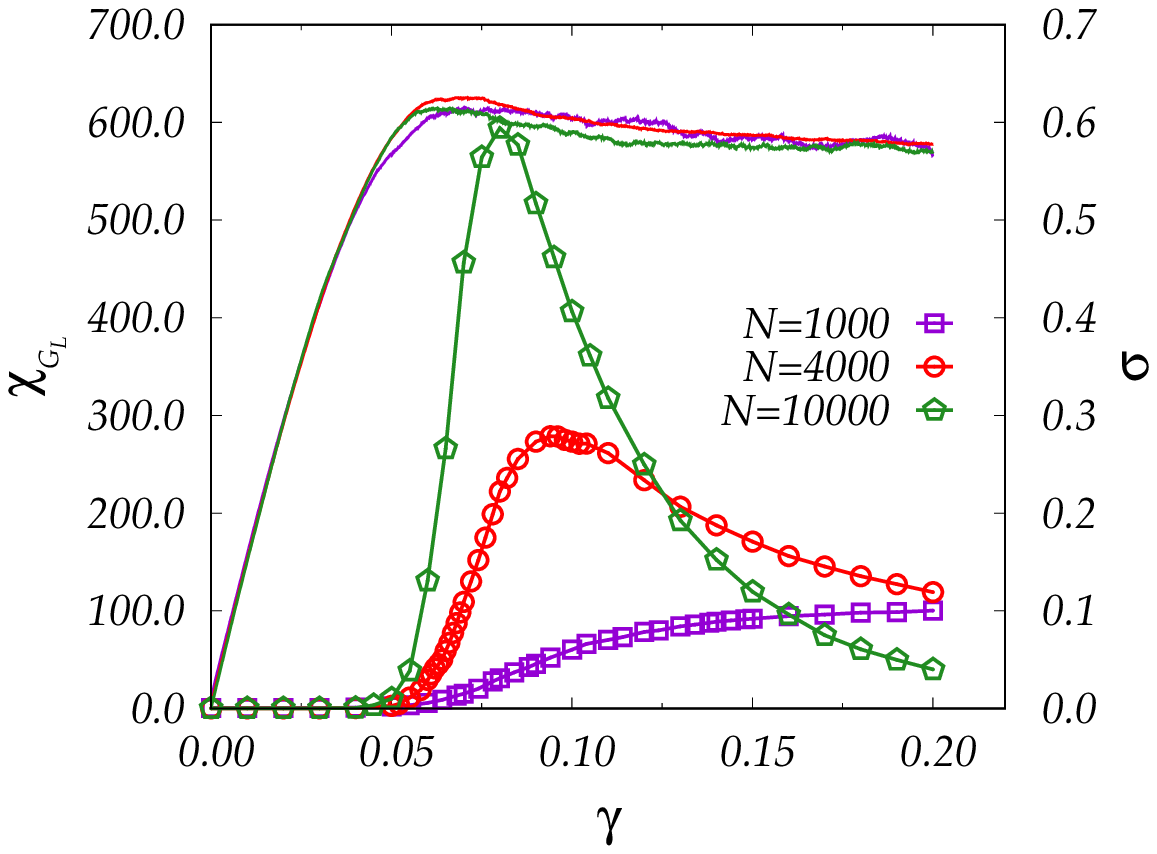}
 \caption{The susceptibilities $\chi_{_{\Gamma_2}}$ (upper panel) and $\chi_{_{G_R}}$ (middle panel) and $\chi_{_{G_L}}$ (lower panel) as a function of $\gamma$ for the three systems sizes available. Superimposed are the stress vs. strain curves for comparison.
 The color code is violet for $N=1000$, red for 4000 and green for 10000.}  \label{sus}
\end{figure}
\subsection{Correlation functions}

 Based on the introductory discussion, we now derive an expression for the correlator associated with the longitudinal mode, from whence one can extract the diverging correlation length associated with the onset of criticality at the spinodal point, and define an associated susceptibility which will shoot up as the spinodal point is approached. The first step is to ``localize" the overlap function and define the $\B r$-dependent quantity
 \begin{equation}
  Q_{12}(\B r) \equiv \sum_{i=1}^N\theta(\ell-|\B r_i^{(1)}-\B r_i^{(2)}|)\delta (\B r -\B r_i^{(1)})\ ,
  \label{defQr}
 \end{equation}
Next, as mentioned above, the expression for the longitudinal correlator in terms of four-replica correlation functions can be found by diagonalization of the correlation matrix $G_{ab;cd}$, (see Appendix \ref{theory}) which is defined as the inverse of the mass matrix $M_{ab;cd}$ of the replicated field theory of the overlap order parameter $Q_{12}$~\cite{98DKT}. The derivation is a matter of standard diagonalization algebra, so we shall not report it here and refer to Appendix \ref{theory} for the details. The expression, employed for example in~\cite{14BB1,14BB2} in the case of a model with spins on a lattice, reads for athermal systems
 \begin{equation}
  G_L(\boldsymbol{r}) = 2G_R(\boldsymbol{r}) - \Gamma_2(\boldsymbol{r}),
 \end{equation}
 with the definitions
 \begin{eqnarray}
 G_R(\boldsymbol{r}) &\equiv& \overline{\left<Q_{12}(r)Q_{12}(0)\right>} - 2\overline{\left<Q_{12}(r)Q_{13}(0)\right>}\\
 &&+ \overline{ \left<Q_{12}(r)\right>\left<Q_{34}(0)\right>},\nonumber\\
 \Gamma_2(\boldsymbol{r}) &\equiv& \overline{\left<Q_{12}(\boldsymbol{r})Q_{12}(0)\right> - \left<Q_{12}(\boldsymbol{r})\right>\left<Q_{12}(0)\right>}\ .
 \end{eqnarray}
We reiterate that angular brackets denote a patch average and $\overline{(\bullet)}$ indicates an average over different patches. The quantity $G_R(\boldsymbol{r})$ is the correlation function of the replicon mode~\cite{98DKT} and $\Gamma_2(\boldsymbol{r})$ is just the garden-variety four-point correlator.

 Using these definitions and taking Eq. \eqref{defQr} into account, the quantities we compute in numerical simulation, before taking the ensemble average, are (see Appendix \ref{theory} and Ref.~\cite{16BCJPSZ}):
 \begin{widetext}
 \begin{equation}
 \tilde \Gamma_2(\B r) = \frac{ \sum_{i\neq j}(u^{(12)}_i-Q_{12}) (u^{(12)}_j-Q_{12})\delta(\boldsymbol{r}-(\boldsymbol{r}_{i}^{(1)}-\boldsymbol{r}_{j}^{(1)})) }{ \sum_{i\neq j}\delta(\boldsymbol{r}-(\boldsymbol{r}_{i}^{(1)}-\boldsymbol{r}_{j}^{(1)})) }
 \end{equation}
 and
 \begin{equation}
  \tilde G_R(\B r)  = \frac{ \sum_{i\neq j}[u^{(12)}_iu^{(12)}_j - 2u^{(12)}_iu_j^{(13)} + Q_{12} ~Q_{34}]\delta(\boldsymbol{r}-(\boldsymbol{r}_{i}^{(1)}-\boldsymbol{r}_{j}^{(1)})) }{ \sum_{i\neq j}\delta(\boldsymbol{r}-(\boldsymbol{r}_{i}^{(1)}-\boldsymbol{r}_{j}^{(1)})) }.
 \end{equation}
  \end{widetext}
 with
 \begin{equation}
  u^{(12)}_i \equiv \theta(\ell-|\boldsymbol{r}_i^{(1)}-\boldsymbol{r}_i^{(2)}|) \ .
 \end{equation}
These four-replica objects can be computed for any quadruplet of distinct replicas. The ensemble averaged correlation functions are simply obtained as $\Gamma_2\equiv \overline{\tilde \Gamma^{(12)}_2}$ and $G_R\equiv \overline{\tilde G^{(12)}_R}$, and cf. Appendix \ref{theory} for a proof. We stress that one must keep the full space dependence of the correlators in the definitions above, as the shear strain breaks the rotational symmetry of the glass samples and so the correlators are not just functions of a distance $r$.
\section{Numerical Results}
\label{numerics}

The three correlation functions discussed in the previous section were computed
{within the same numerical framework as} discussed in Subsect.~\ref{transition}. Typical results
are shown in Fig.~\ref{correlations}, here at $\gamma_S=0.09405$. One notices the obvious fact that the correlation functions reflect the breaking of isotropy that is caused by the strain. In fact the {spatial} structure of the correlation
function is quadrupolar, precisely indicating where shear bands are bound to appear in simple strain: the $x$ and the $y$ axes are in $45^\circ$ to the principal axis of the stress~\cite{13JGPS}.

To demonstrate the strain dependence of the correlators
we consider first the susceptibilities $\chi_{_{G_L}}, \chi_{_{G_R}}$ and $\chi_{_{\Gamma_2}}$ that can be obtained from the correlators. For example
\begin{equation}
 \chi_{_{G_L}}(\gamma) \equiv \int d^2 x\ G_L(x,y;\gamma) \ .
\end{equation}

 In Fig.~\ref{sus} upper panel we show the susceptibility $\chi_{_{\Gamma_2}}$ as a function of $\gamma$ for the three system sizes at our disposal. Superimposed are the stress vs. strain curves obtained by averaging
 the individual curves over all the available configurations and glass samples. One sees very clearly the singularity that
 develops near the spinodal point as a function of the system size. In the middle and lower panel of the same figure we show the other two susceptibilities $\chi_{_{G_R}}$ and $\chi_{_{G_L}}$ as a function of the strain $\gamma$, again with the stress-strain curve superimposed for comparison. As we expected, the susceptibilities show a distinct peak at the spinodal point $\gamma_S$ where criticality is reached.

 The scaling of the peak of the susceptibility $\chi_{_{\Gamma_2}}$ with the system size is expected
 to mirror the scaling of the response as written in Eq.~(\ref{SN}), at least if standard fluctuation-dissipation
 theorems should apply to the present problem. Indeed, plotting the maximal values of $-\chi_{_{\Gamma_2}}$ as
 a function of $N$ in a log-log plot, cf. Fig.~\ref{-chi} we find that the maxima $\chi^{\rm max}_{_{\Gamma_2}}$ scale like $\sqrt{N}$ as expected.
  \begin{figure}
\includegraphics[width=0.5\textwidth]{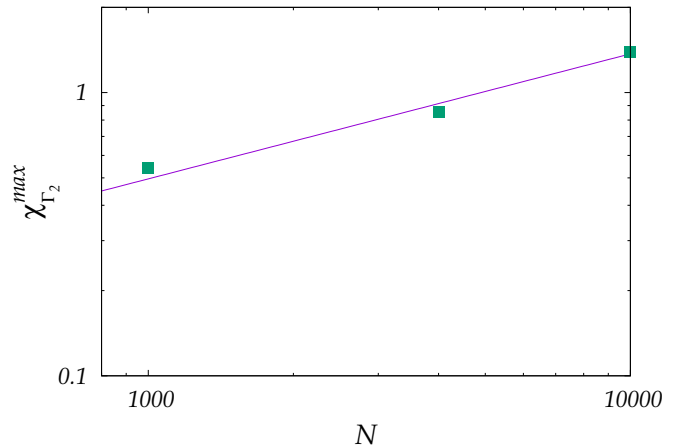}
 \caption{The system size dependence of the maxima $\chi^{\rm max}_{_{\Gamma_2}}$ indicating
 a dependence on $\sqrt{N}$ {consistently with} the results in Fig.~\ref{sharpness} }
 \label{-chi}
 \end{figure}

 More detailed information is provided by the full dependence of the correlators on their arguments.
 To see most clearly the change in the correlators as the spinodal point is approached, we
consider for example the one-dimensional function $G_R(x=0,y;\gamma)$, shown for $N=4000$ in Fig.~\ref{cut}.
\begin{figure}
\includegraphics[width=0.5\textwidth]{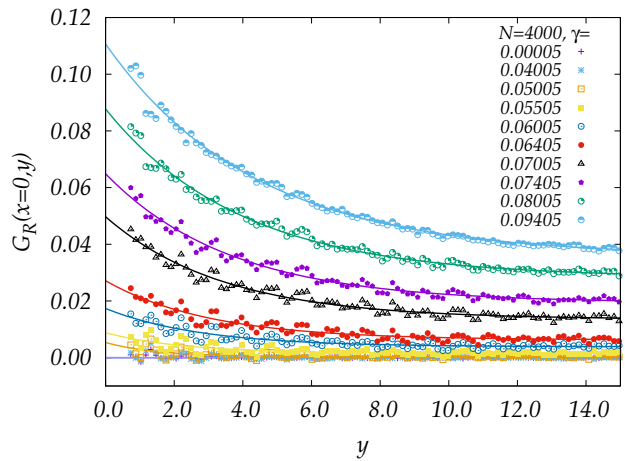}
 \caption{The function $G_R(x=0,y;\gamma)$ for various values of $\gamma$ from $5\times 10^{-5}$ to 0.09405.  Note the increase in the
 over all amplitude of the correlator as well as the increase in the correlation length. The lines through
 the data are the fit function Eq.~(\ref{fit}).}
 \label{cut}
\end{figure}
\begin{figure}
\includegraphics[width=0.4\textwidth]{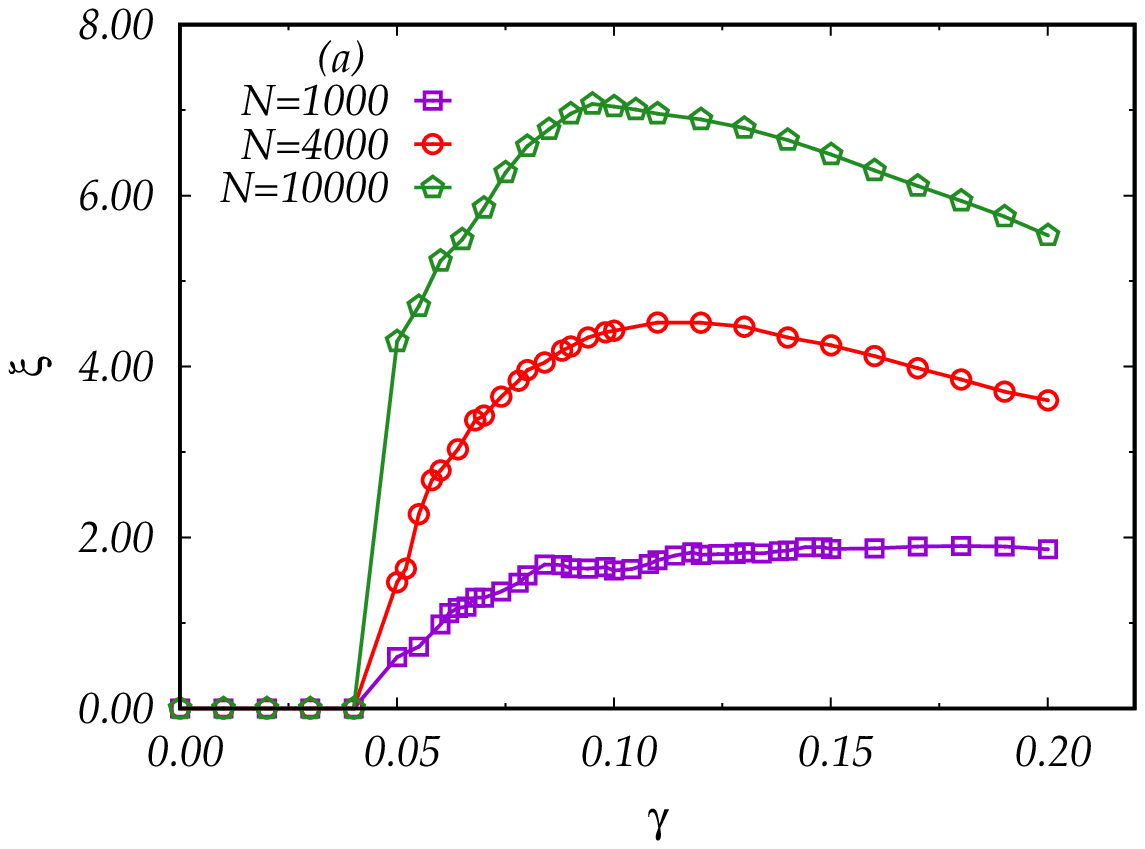}
\includegraphics[width=0.4\textwidth]{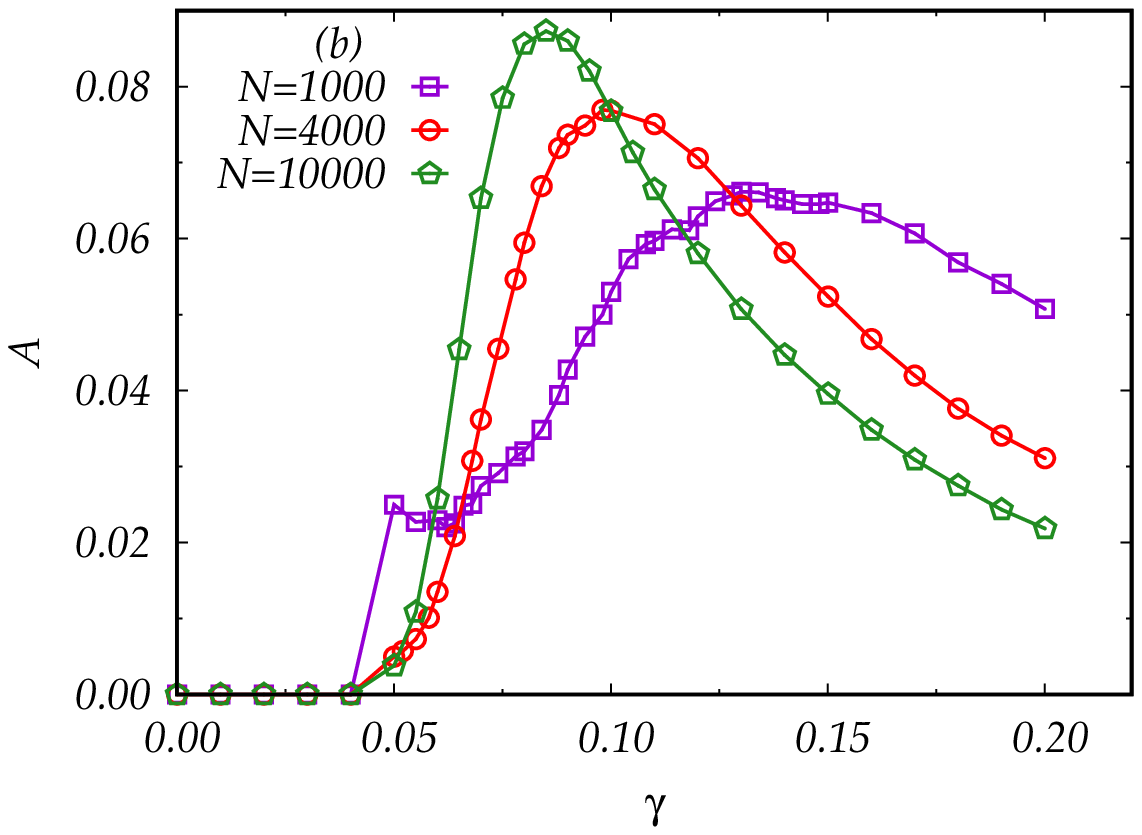}
\includegraphics[width=0.4\textwidth]{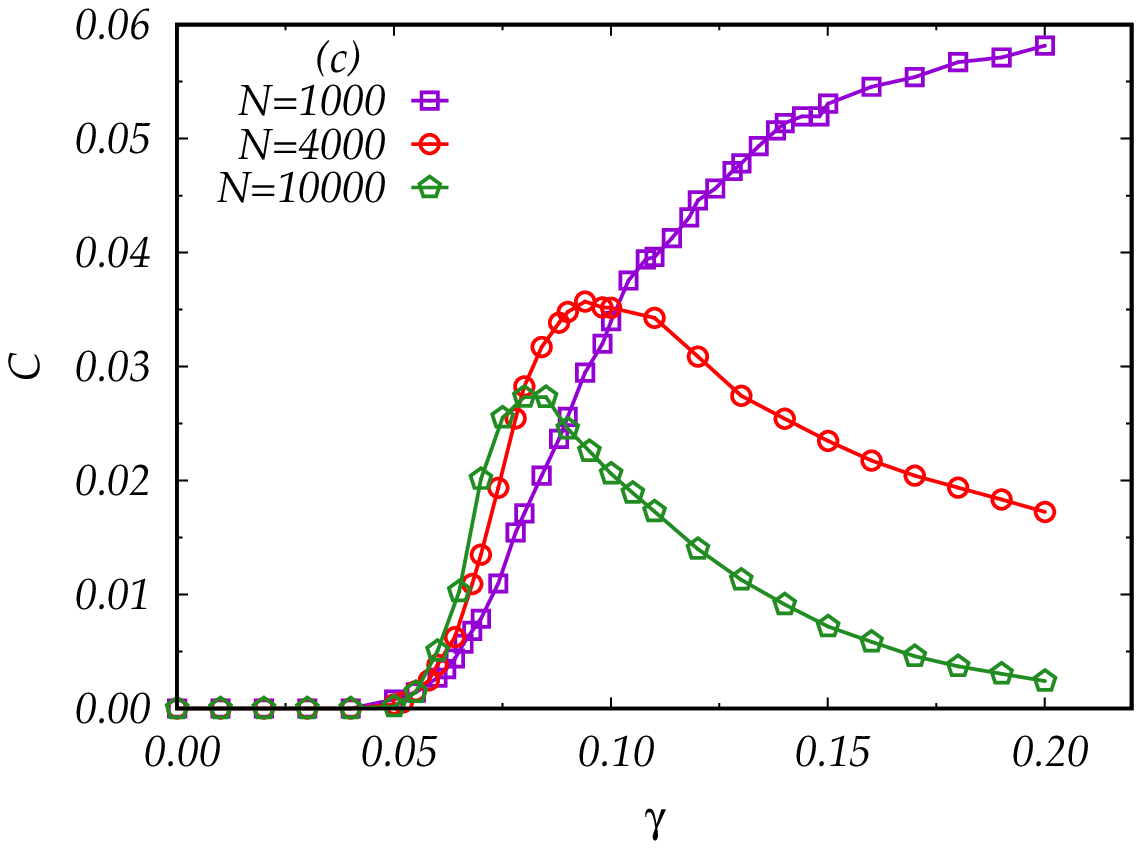}
 \caption{The $\gamma$ dependence of the correlation length $\xi(\gamma)$, the amplitude $A(\gamma)$ and the constant $C(\gamma)$ in the
 best fit to the function $G_R(x=0,y;\gamma)$, cf. Eq.~\ref{fit}. }
 \label{results}
\end{figure}
We note that the correlator changes both in amplitude and in extent when we approach the critical point.
To quantify these changes we fit a 3-parameter function to $G_R(x=0,y)$ in the form
\begin{equation}
G_R(x=0,y;\gamma)\approx C+ A \exp(-y/\xi) \ ,
\label{fit}
\end{equation}
where all the fitting coefficients are functions of $\gamma$.
In Fig.~\ref{results} we present the $\gamma$ dependence of the amplitude $A(\gamma)$, the constant $C$ and the correlation length $\xi(\gamma)$.
 \begin{figure}
 \includegraphics[width=0.40\textwidth]{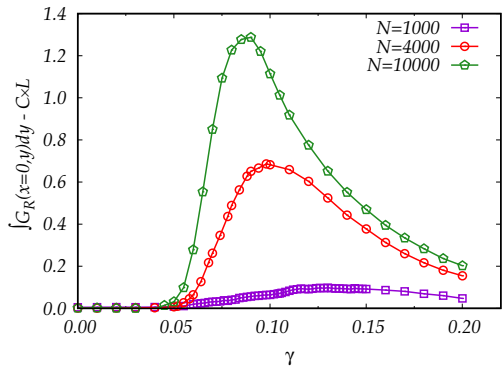}
 \caption{The difference between $\int dy\ G_R(x=0,y)$ and $C\times L$}
\label{Ceffect}
\end{figure}

An interesting observation concerns the constant $C$ used in the fit Eq.~(\ref{fit}). This constant
is also sensitive to the approach of the criticality, cf. the lower panel in Fig.~\ref{results}. One could worry
that integrating this constant over $y$ could contribute to the divergence of the susceptibilities. In fact
the rise in $C$ near the spinodal point goes down with the system size and its contribution to the integral
is reduced as well, as can be seen in Fig.~\ref{Ceffect} which presents the integral $\int dy G_R(x=0,y)$ from
which $C\times L$ is subtracted.
The conclusion is that indeed the contribution of $C$ goes down also when integrated over the system
size, showing that the main contribution to the divergence of the susceptibility is from the divergence
of the correlation length.
It is interesting to notice that the constant $C$ decreases with the system size, presumably becoming irrelevant
in the thermodynamic limit. The amplitude $A$ is still increasing with the system size, and it is difficult
to assert whether it converges or not. On the other hand we can safely conclude that the data present a strong evidence for the increase in the correlation length. This conclusion is substantiated below using
the correlation function $\Gamma_2(x,y)$. Before doing so
we need to discuss the fitting procedure for the correlation function $G_R(x=0,y)$. In Fig.~\ref{full}
we show the full results for this correlation function for all the available values of $\gamma$ and for two larger systems sizes at our disposal. One sees that the exponential decay that is used for the fit is only
reliable up to the minima of the functions. The reason for the upward trend is the periodic boundary condition that reflects the correlations. To eliminate this spurious effect we presented in Fig.~\ref{cut} the fit
up to the minimum in the function. One should note however that the distance to the minimum increases
with the system size, presumably diverging in the thermodynamic limit. Thus the fit up to the minimum
allows a faithful estimate of the correlation length $\xi$.
\begin{figure}
 \includegraphics[width=0.4\textwidth]{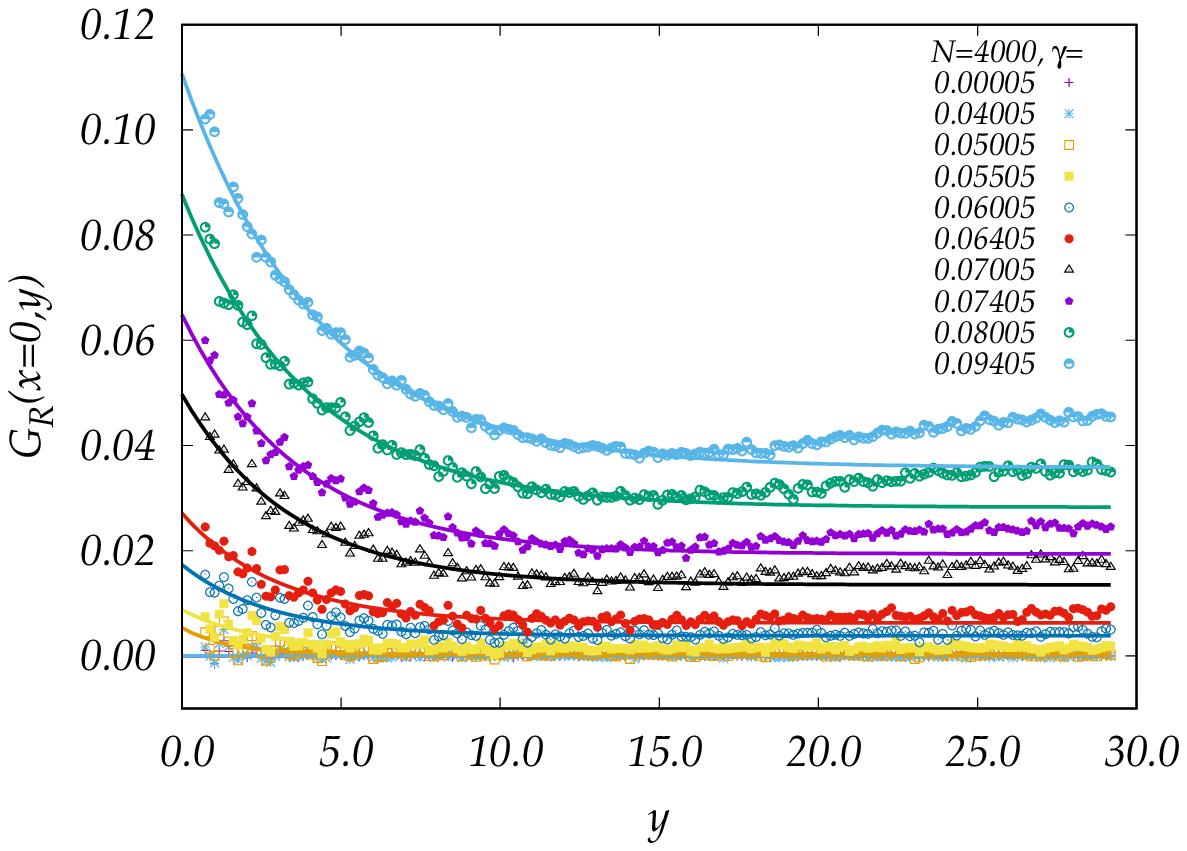}
 \includegraphics[width=0.4\textwidth]{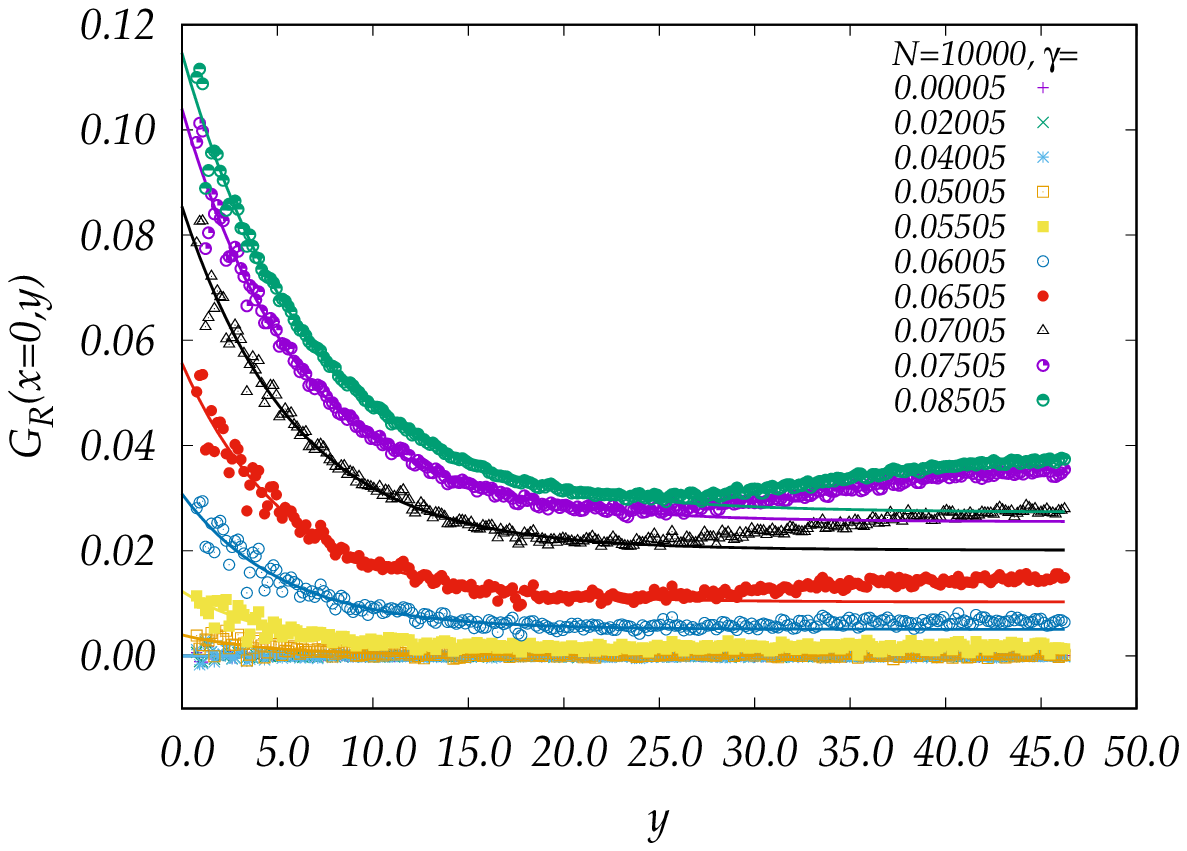}
\caption{The full $y$ dependence of $G_R(x=0,y)$. The region fitted by Eq.~(10) in the paper
is shown. Note that the minimum in the function resides in higher values of $y$ for larger
system sizes.}
\label{full}
\end{figure}

The dependence of $\xi$ on the distance from criticality and on the system size is not easy to
read from Fig.~\ref{results}. In fact a smoother dependence is available from the correlation
function $\Gamma_2(x=0,y)$ and $\Gamma_2(y=0,x)$. An exponential fit similar to Eq.~(\ref{fit}) was
applied to these two projections of $\Gamma_2(x,y)$ and the correlation length $\xi$ was determined as
shown in the upper panel of Fig.~\ref{goodxi}.
\begin{figure}
 \includegraphics[width=0.4\textwidth]{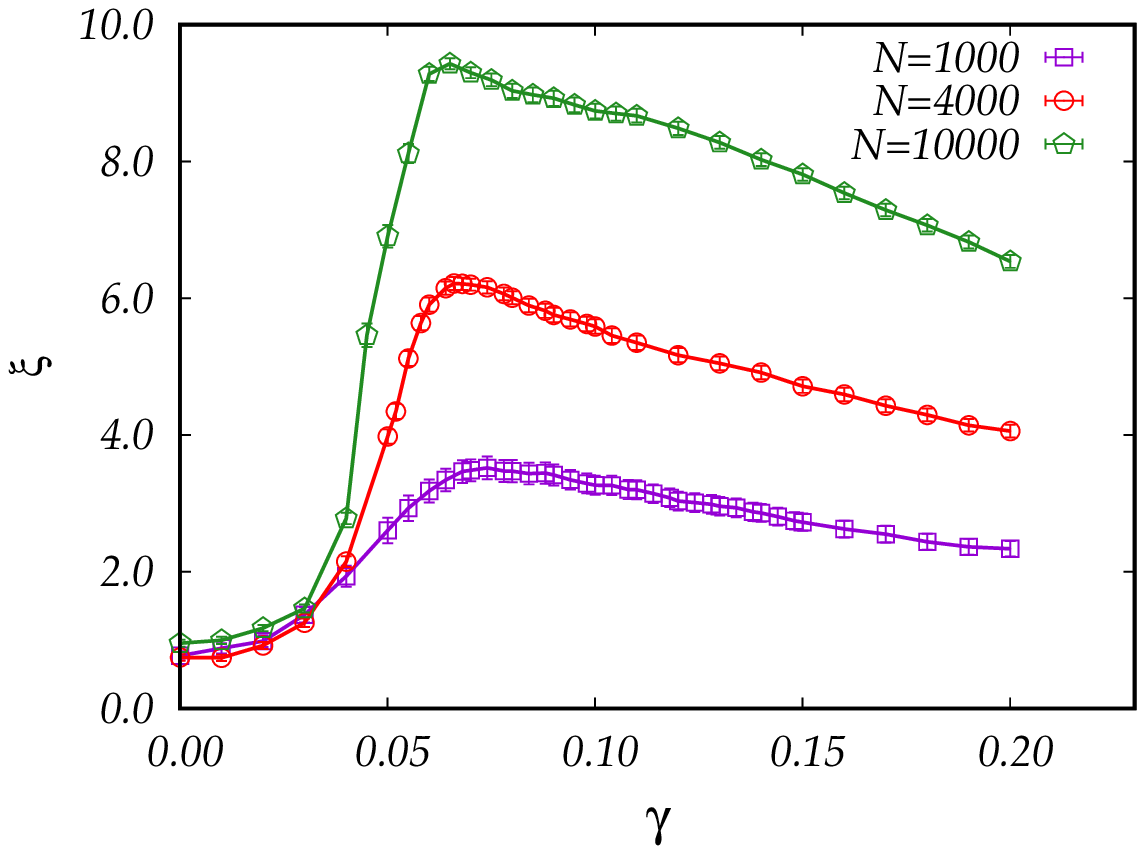}
 \includegraphics[width=0.5\textwidth]{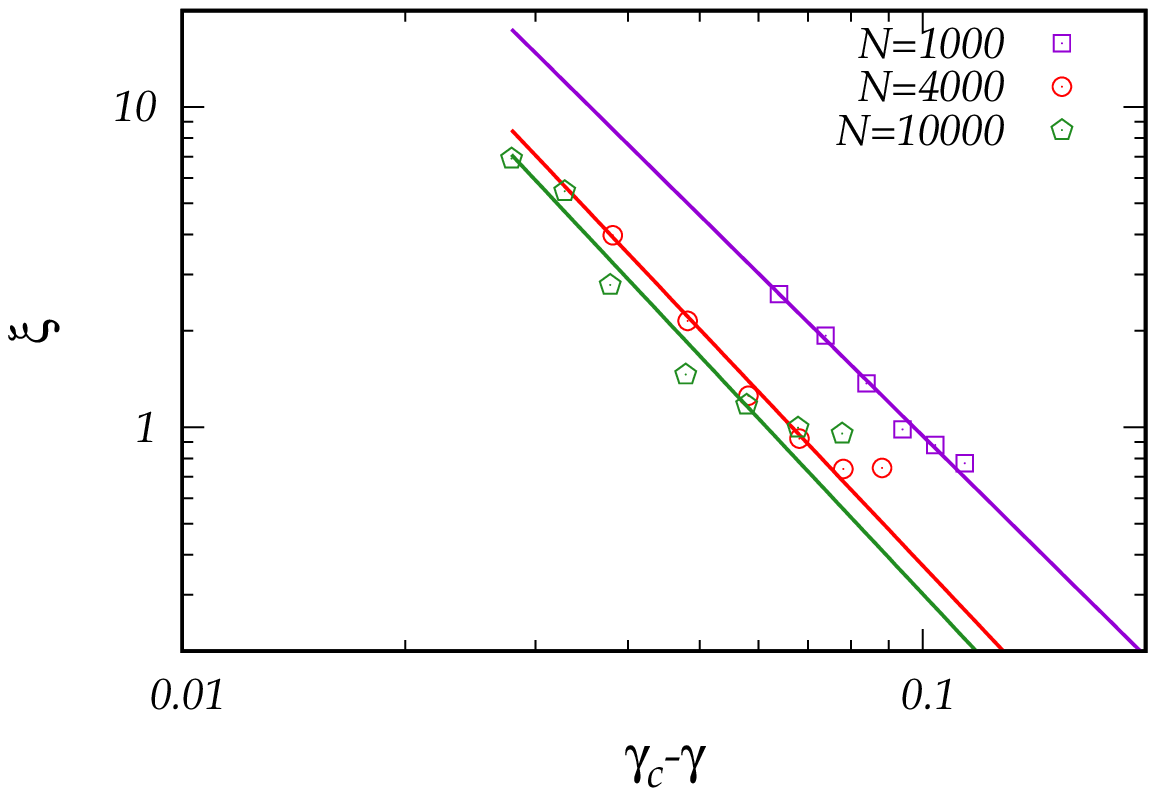}
\caption{Upper panel: the correlation length $\xi$ read from an exponential fit to the
$x$ and $y$ projection of the correlation function $\Gamma_2(x,y)$ for three values of
the system size. Lower panel: the dependence of the correlation length $\xi$ on $\gamma-\gamma_c$.}
\label{goodxi}
\end{figure}
The scaling exhibited in the lower panel of Fig.~\ref{goodxi} is not perfect, but a least
square fit to all the three curves leads to a scaling law in the form
\begin{equation}
\xi \approx (\gamma_c-\gamma)^{-\nu} \ , \quad \nu\approx 2.4\pm 0.35 \ .
\label{xilaw}
\end{equation}
The estimated value of $\nu$ is unusually high. The error bars are significant, and it is quite likely that this result indicates that $\nu=2$, although at the present time we cannot
offer a theoretical basis for this number.

The result Eq.~(\ref{xilaw}) may have important experimental consequences, predicting the length
of micro-shear bands in materials as a function of the distance from criticality. We propose that such
measurements should be carried out, providing a possible direct test of the present ideas.

\section{The effects of finite tempeartures}
\label{T}

The mechanical yield in athermal conditions is an excellent conceptual laboratory for clarifying
the essence of the yield mechanism, but in reality many yielding amorphous solids operate under
thermal conditions, effected by thermal fluctuations. It is therefore interesting and important
to assess the effects of temperature on the findings described above.

To assess the effects of temperature we repeat precisely the same protocol described above
to create a patch of $n$ replica at $T=0$, including the creation of $m$ such patches. The difference
is that presently we warm up all the replica in a given patch to a target temperature. Results will
be reported for target temperatures $T=0.1, 0.2$. While keeping the strain at $\gamma=0$ each configuration
was thermalized by molecular dynamics. {Afterwards,} each configuration was strained by increasing
the strain in steps of $\delta \gamma=2\times 10^{-4}$, allowing the energy to stabilize after each such step
before straining again. Typical averaged strain vs stress curves (with averages computed firstly over a patch
and secondly over all the patches) for a system with $N=10000$ are shown in the upper panel of Fig.~\ref{Tsigvsgam}.
\begin{figure}
\includegraphics[width=0.5\textwidth]{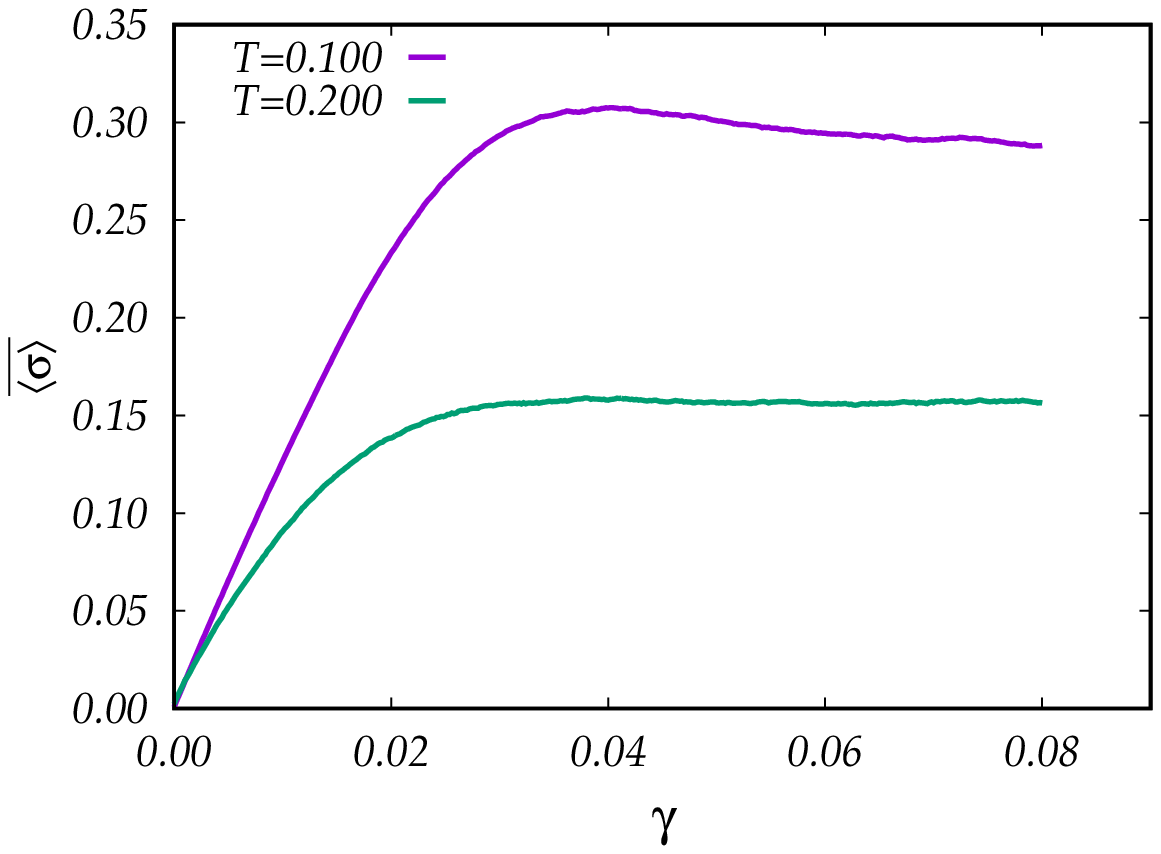}
\includegraphics[width=0.5\textwidth]{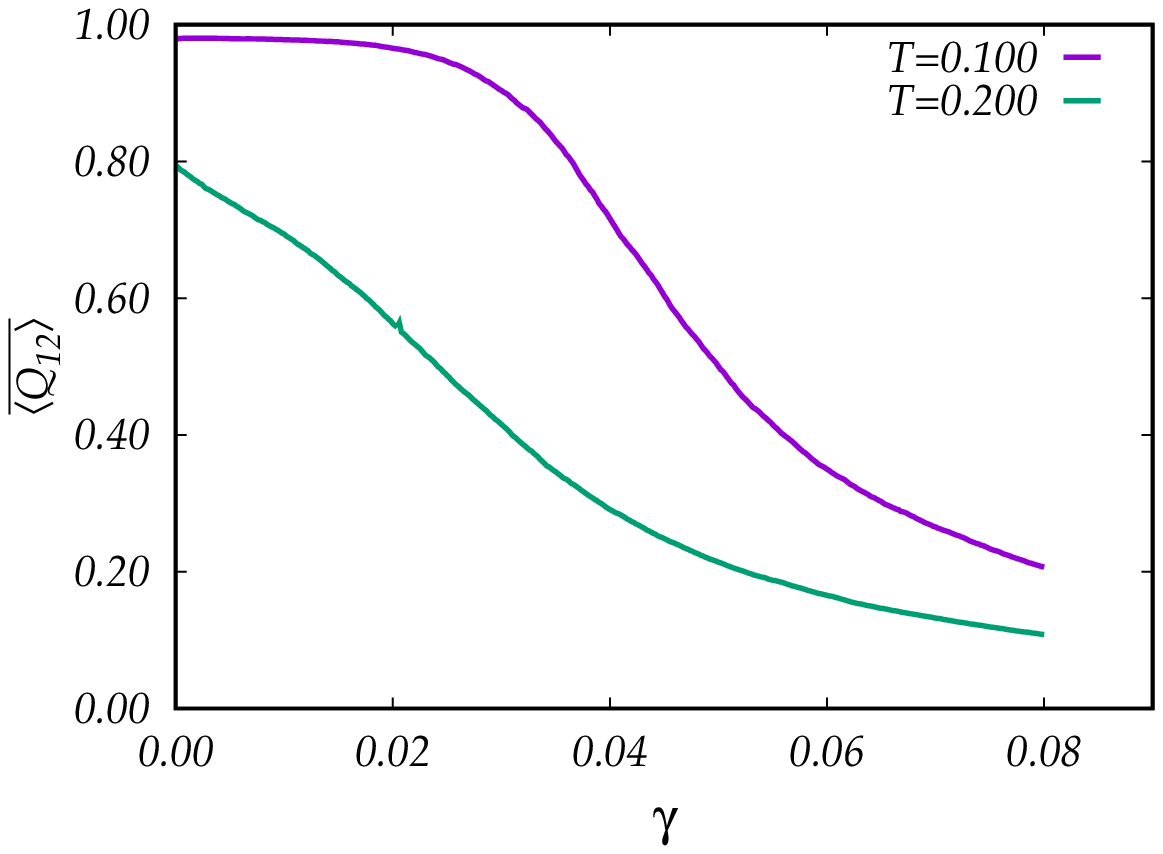}
\caption{Upper panel: average stress vs. strain in quasi-static straining at finite temperature.
The average is computed over all the configurations in a patch and then over the patches. Lower
panel: the averaged overlap order parameter computed at the designated temperatures as a function
of the strain $\gamma$. Here $N=10000$.}
\label{Tsigvsgam}
\end{figure}
We see that at the lower temperature $T=0.1$ there is still a stress peak before the yield, but
at the higher temperature $T=0.2$ the stress peak no longer exists and the stress reaches the
flow steady state stress quite monotonically. {At} both temperatures the steady state is attained at
lower values of the strain than at $T=0$. Computing the average overlap order parameter $\overline{\langle Q_{12}\rangle}$
(cf. lower panel of Fig.~\ref{Tsigvsgam}) we observe a corresponding behavior. At $T=0.1$ a remnant phase
transition is still observable, with the order parameter {still} falling {somewhat} sharply after $\gamma\approx 0.04$. At
$T=0.2$ there is no longer a sharp decrease, but rather a smooth decline of $\overline{\langle Q_{12}\rangle}$ as a function
of $\gamma$. It is obvious that temperature fluctuations at $T=0.2$ are sufficient to destroy the
spinodal characteristics of our phase transition.

The same conclusion is drawn from examining the pdf of the order parameter. In Fig.~\ref{Tpdf} we show
the function $\overline{P_\gamma(Q_{12},T)}$ for three temperatures $T=0, 0.1, 0.2$ for a system
with $N=10000$.
\begin{figure}
\includegraphics[width=0.4\textwidth]{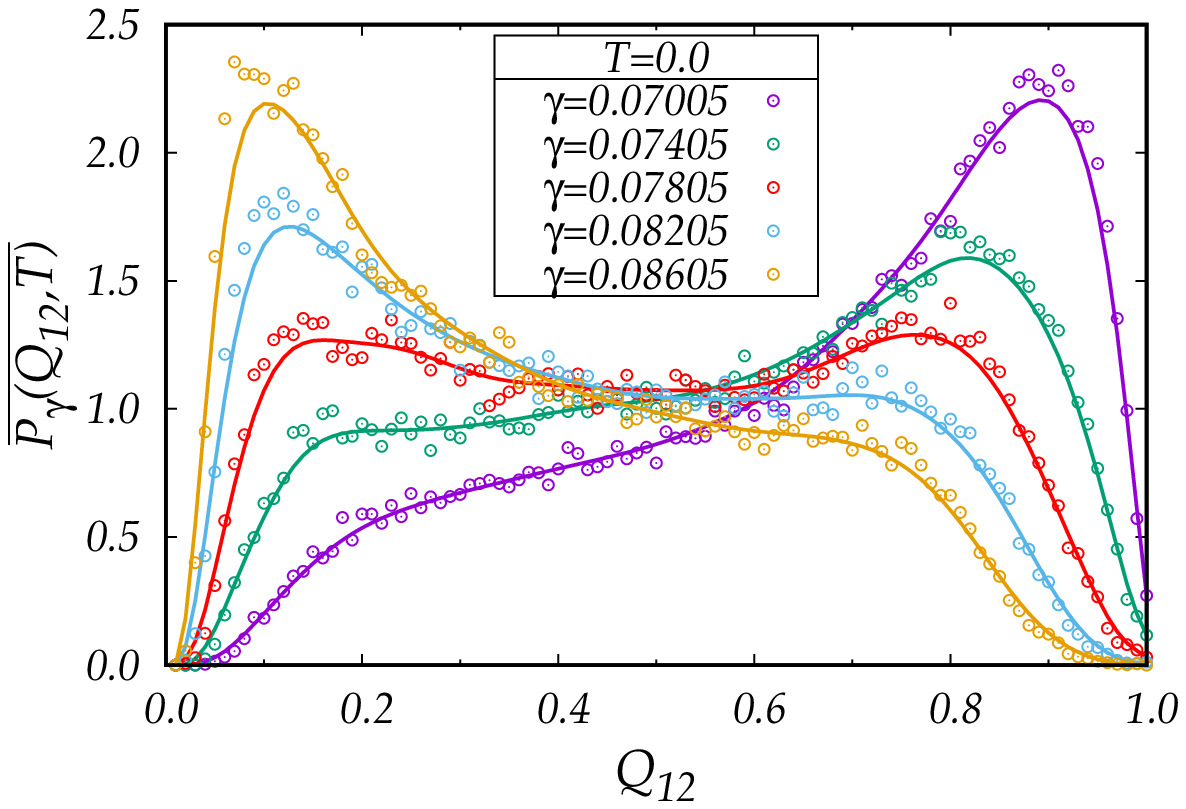}
\includegraphics[width=0.4\textwidth]{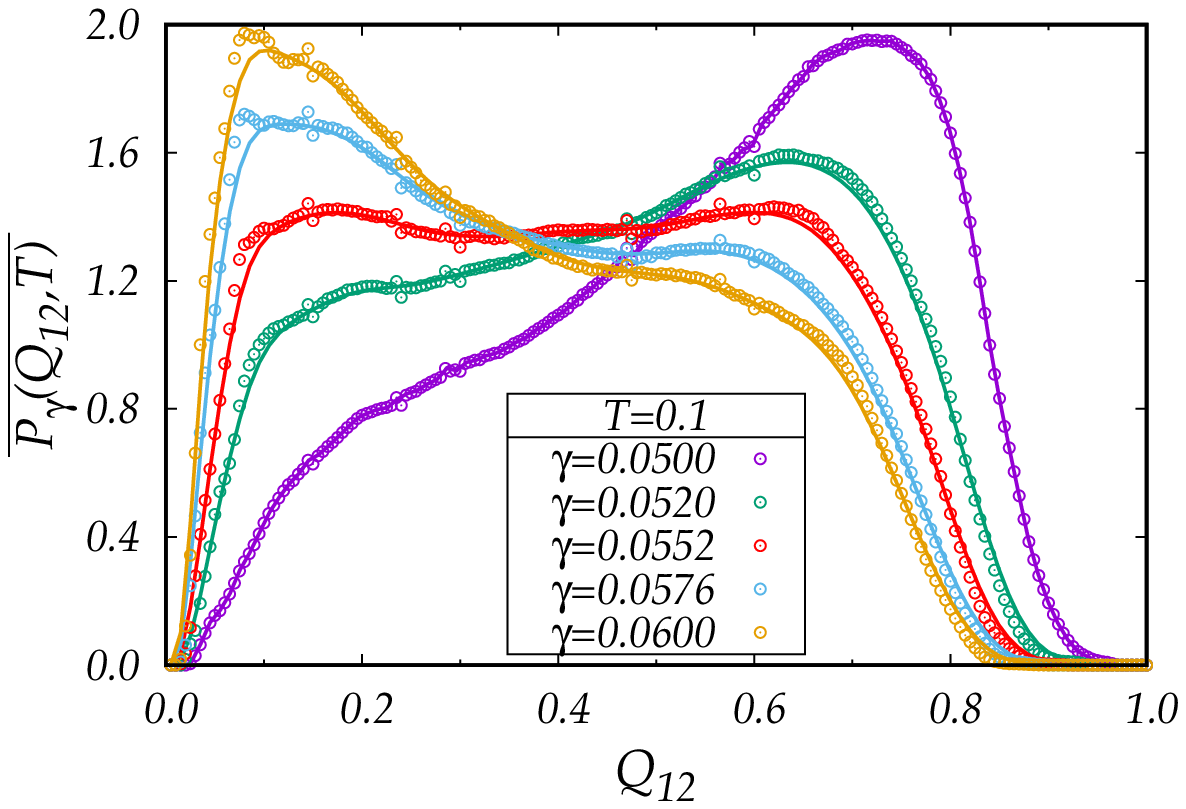}
\includegraphics[width=0.4\textwidth]{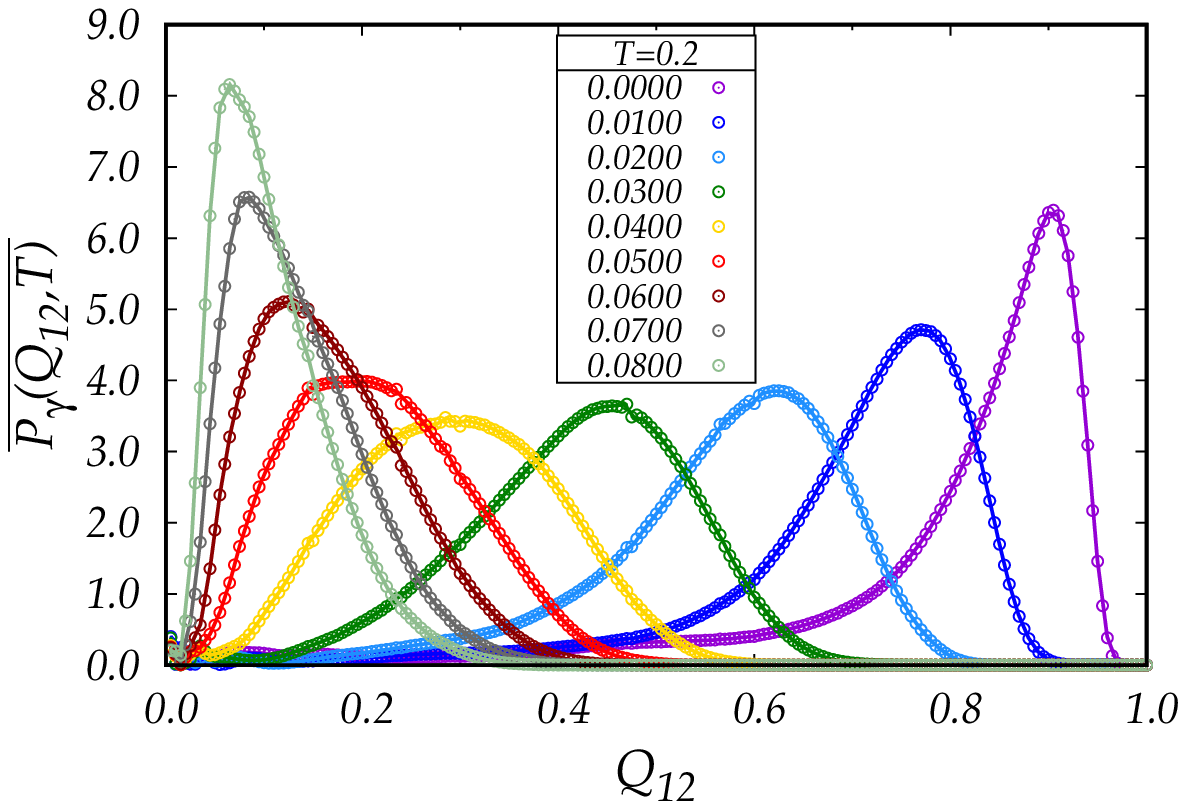}
\caption{The dependence of $\overline{P_\gamma(Q_{12},T)}$ on $\gamma$ for three
different temperatures $T=0,0.1,0.2$.}
\label{Tpdf}
\end{figure}
While the phase transition is observed nice and clear at $T=0$, {and} it still remains observable
at $T=0.1$, it changes to a smooth migration of the single peak of $\overline{P_\gamma(Q_{12},T)}$
from high to low values of $Q_{12}$ when $\gamma$ is increased. We lose completely the double hump
structure which underlies the spinodal criticality.

It is important to stress that with the loss of the spinodal criticality we also lose the qualitative
distinction between the pre-yield and post-yield statistics of the energy drops as shown in Eqs.~(\ref{UN0}) and
(\ref{UN}). Observing a stress vs. strain curve for a single realization one finds the same statistics
of energy and stress fluctuations before and after the yield, since it is dominated now by temperature fluctuations
rather than by mechanical instabilities. The sharp appearance of system spanning
events with the yield phenomenon is caused by the spinodal criticality as explained in this paper.
Once this gets destabilized by temperature fluctuations there is no increase in correlation length
and we remain only with standard temperature fluctuations.

\section{Summary and Conclusions}
\label{summary}
In summary, we have presented evidence that the scale free yielding transition in amorphous solids is
governed by a spinodal point with disorder. The associated {correlation} length is exhibited by suitable four-point correlators whose expression can be obtained from replica theory. The full implications of the theory pertain to an athermal setting, and the full fledged criticality is destroyed by thermal fluctuations~\cite{16SCH}. In athermal conditions the transition becomes ever sharper with increasing the system size. We have found that the
range of strain values over which the transition takes place goes to zero like $1/\sqrt{N}$. The correlation
length $\xi$ appears to diverge {following} a scaling law, cf. Eq.~(\ref{xilaw}). We have commented above that this prediction may be tested experimentally by examining the lengths of micro shear bands as a function of the strain or the
stress while approaching mechanical collapse.

For sufficiently high temperatures the system will generally be able to escape through thermal activation from the high-$Q_{12}$ minimum before this has a chance to flatten and the relative susceptibility to diverge. However, since the nucleation time is expected to be fairly long, one should anyway be able to observe \emph{transient} shear-bands/heterogeneities, as long as the temperature is low enough that nucleation does not take place until the system is close to the spinodal, which, interestingly, is precisely the behavior of transient shear bands as reported in~\cite{16SCH}.

Finally we need to touch upon the notion of `self organized criticality'. Not being quite sure what it means, we propose that
it refers to the fact that after yield the system remains critical in the sense that the yielded configurations are still maintained
at the yield stress, and are therefore marginal; any increase in strain will cause a repeat of the phenomena discussed
above. To see this we need to recreate a patch of configurations that are closed to a given yielded configuration and examine what
happens upon further straining of such a patch. To create a patch of configurations having almost same value of stress $\sigma\approx \sigma_{_{\rm Y}}$, we take a single post-yield configuration, and apply random displacements  of randomly selected particles (keeping the displacements infinitesimal, so that the overlap function $Q_{12}$ remains close to unity).  We then perform conjugate-gradient minimization to return each configuration to an athermal mechanical equilibrium (T=0). The obtained configurations are close to the selected post-yield configuration and have almost the same stress. This method is repeated to generate as many configurations as we need that belong to an approximately iso-stressed  patch. Finally we strain al the configurations in the patch and  compute the strain dependent order parameter as explained above.
The result of this analysis is displayed in Fig.~\ref{after}.
\begin{figure}
 \includegraphics[width=0.4\textwidth]{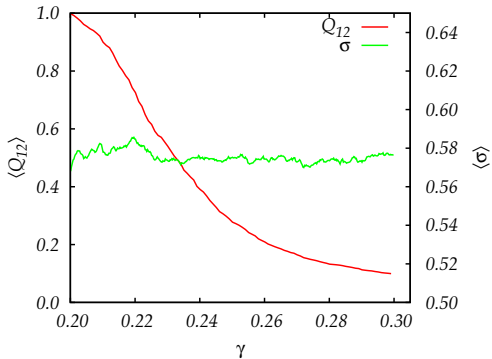}
 \caption{The dependence of the order parameter (averaged over 10 patches of stressed configurations) on the strain
 in athermal conditions. We see that now the order parameter decreases immediately and continuously, without any
 sign of a phase transition.}
 \label{after}
 \end{figure}
 We see that the phase transition is now eliminated, the order parameter decreases smoothly and without a sharp
 decline at any specific value of $\gamma$. The reason is of course that each patch contains many marginal
 configurations that yield again and again in the strain controlled experiment. We reiterate that in stress controlled
 experiments the system collapses anyway when the yield-stress is exceeded. The conclusion is that the criticality is not
 at all self-organized, it is caused due to the mechanical straining by the external agent; the system is driven
 to a marginal state and is maintained there by continuing to strain the system.

In the future one needs to examine further the universality of the proposed scenario and of the scaling laws
found in this paper, examining different amorphous systems and different space dimensionalities. Another interesting future path is the study of the mechanical yield in frictional aggregates. It is not known
at the present point in time whether these {systems} fall in the same universality class or whether they might
exhibit totally different behavior.

\section{Acknowledgements}
 We thank Giorgio Parisi for inspiring discussions. We acknowledge interesting exchanges with Giulio Biroli regarding the effect of disorder on the width of the transition. This work has been supported
in part by the Minerva foundation with funding from the Federal German Ministry for Education and Research, and
by the Israel Science Foundation (Israel Singapore Program).

\appendix
\section{The longitudinal correlation function}
\label{theory}
Let us start from the expression of the free energy of a glass state, prepared by equilibrating a generic glass former down to a glass transition temperature $T_g$ where it can still be equilibrated, and then quenching it out of equilibrium to a given temperature $T<T_g$. Such a free energy was first defined in~\cite{FP95} in the context of spin-glass physics. Its definition in the case of structural glasses, and its computation in the particular case of hard spheres were first discussed in~\cite{15RUYZ}. The definition, in the case of a generic glass former made of $N$ particles is based on comparing two configurations $X^a$ and $X^b$ of the same glass. Here
\begin{equation}
X^a \equiv \{\B r_i^a\}_{i=1}^N \ , \quad X^b \equiv \{\B r^b_i\}_{i=1}^N \ ,
\end{equation}
where the labeling $\B r_i$ refers to the position of the same particle $i$ in the two different configurations.
For a generic interaction potential $V(X)$ the definition of the free energy is
\begin{eqnarray}
f[T,T_g] &\equiv& -\frac{1}{\beta N} \int dX_0\ \frac{e^{-\beta_g V(X_0)}}{Z_g}\nonumber\\
 &\times&\log\left[\int dX_1\ e^{\beta V(X_1)}\delta(q^*_r - Q_{01})\right].
\label{eq:FP}
\end{eqnarray}
where $\beta_g=1/(k_B T_g)$, $\beta=1/(k_B T)$ and $q^*_r$ is the value of $q_r \neq 0$ whereupon the free energy attains a local minimum~\cite{15RUYZ}.
The overlap function $Q_{01}$ for any two configuration, say $a$ and $b$ is~\cite{16JPRS}
\begin{equation}
 Q_{ab} = \frac{1}{N}\sum_{i=1}^N \theta(\ell-|\boldsymbol{r}^a_i - \boldsymbol{r}^b_i|).
\end{equation}
Here $\ell$ is a coarse graining parameter (in~\cite{16JPRS}, $\ell
\simeq 0.3$ in Lennard-Jones units).
The idea is to consider the free energy at temperature $T$ of the glass former, which is \emph{constrained} to stay close to an amorphous configuration $X_0$ which is selected from the equilibrium ensemble, using the canonical distribution when the glass is still at equilibrium at $T_g$.

The properties and computation of the free energy \eqref{eq:FP} are discussed extensively in~\cite{15RUYZ,16RU}, so we refer the interested reader to those works. The explicit analytic computation is accomplished in the
mean field approximation. In our paper we use the results far from the mean field limit, but we ascertain that the relevant correlation functions that are fleshed out in the mean field calculation are the relevant ones also in the general case. Of course, critical exponents can differ. In the sequel we sketch how from this mean-field theory in terms of an overlap order parameter $Q_{ab}$ one can extract the definitions of the correlation functions that are expected to show critical behavior.

The outermost integral in the Eq.~\eqref{eq:FP} can be computed with the replica trick,
\begin{equation}
 f[T,T_g] = \lim_{s\to 0} \partial_s \Phi[T,T_g;s],
\end{equation}
where $\Phi$ is defined as
\begin{widetext}
\begin{equation}
 \Phi[T,T_g;s] =  -\frac{1}{\beta N}\log \int dX_0\ dX_1 \cdots dX_s e^{-\beta_g V(X_0)} e^{- \beta V(X_1)}\delta(q-Q_{01})\cdots e^{- \beta V(X^s)}\delta(q-Q_{0s}),
\end{equation}
\end{widetext}
so we are considering $s$ replicas of the $X$ configuration. In infinite dimensions for the case of hard spheres it was shown~\cite{KPZ12} that the functional defined above can be written as
\begin{equation}
 \Phi = -\frac{1}{\beta N} \int \mathcal{D} Q_{ab}\ e^{-d S(Q_{ab})} \ .
\end{equation}
Here $\mathcal{D}Q_{ab}$ denotes an integration measure over all the distinct $Q_{ab}$s,
\begin{equation}
 \mathcal{D}Q_{ab} \equiv \prod_{a<b}^{0,s} dQ_{ab},
\end{equation}
and $d$ is the number of spatial dimensions. The functional $S(Q_{ab})$ is referred to as the ``replica action". In the mean-field limit $d\to\infty$, the integral above can be computed via the saddle point method~\cite{BenderAdvancedMethods}, which means that one must consider the optimum points in $Q_{ab}$ of the replica action $S(Q_{ab})$. This means that $S(Q_{ab})$ plays the role of a \emph{Gibbs free energy}, i.e. the free energy for fixed order parameter. An illustrative example is the case of a Curie-Weiss model (mean-field ferromagnet) wherein, for the Helmholtz free energy $F$ in zero magnetic field, one has~\cite{replicanotes}
\begin{equation}
 F(h=0,T) = \min_{m}G(m,T)
\end{equation}
where $G(m,T)$ is indeed the Gibbs free energy for fixed magnetization $m$. The minimization equation for $G$ is then the celebrated equation for the spontaneous magnetization
\begin{equation}
 \frac{\partial G}{\partial m} = 0 \Longrightarrow m = \tanh(\beta m)
\end{equation}
and the ferromagnetic phase transition takes place when the paramagnetic, $m=0$ minimum of $G$ flattens and splits in two degenerate minima with $m\neq 0$, which implies that at the critical temperature $\frac{\partial^2 G}{\partial m^2} = 0$. The derivation of the $S(Q_{ab})$ action in the case of mean-field hard spheres can be found in~\cite{KPZ12}.

In the present case the $f[T,T_g]$ plays the role of the Helmholtz free energy $F$ and the $S(Q_{ab})$ of the Gibbs free energy $G$. With this analogy, one can understand how the critical properties of glass states are related to the matrix of second derivatives of the replica action $S(Q_{ab})$,
\begin{equation}
 M_{ab;cd} \equiv \frac{\partial^2 S}{\partial Q_{a<b}\partial Q_{c<d}},\qquad a,b,c,d \in [1,s]
\end{equation}
in the limit $s \to 0$ (we stress that $X_0$ is not involved in this definition). The inverse $G_{ab;cd}$ of the tensor $M$, defined as
\begin{equation}
\sum_{e\neq f} M_{ab;ef}G_{ef;cd} = \frac{\delta_{13}\delta_{bd} + \delta_{ad}\delta_{bc}}{2}
\end{equation}
is then the covariance matrix of the mean field theory
\begin{equation}
 G_{ab;cd} = \overline{\left<(Q_{ab}-\left<Q_{ab}\right>)(Q_{cd}-\left<Q_{cd}\right>)\right>},
\end{equation}
wherein the angled brackets denote the thermal average restricted to a single glass sample at temperature $T$ (that is over the canonical distribution of the $X_1$ configuration in the \eqref{eq:FP}), and the overbar denotes the average over all possible glass samples selected at $T_g$ (that is over the canonical distribution of the $X_0$ configuration in the \eqref{eq:FP}). This covariance tensor encodes the critical fluctuations of the system near the critical points whereupon the tensor $M_{ab;cd}$ develops a zero mode.

Let us now assume that the glass state under study is a single minimum of the free-energy landscape of the system wherein all replicas from $1$ to $s$ can move ergodically, this means that the replicas are all equivalent and the matrix $Q_{ab}$ must then be invariant by any replica permutation, an hypothesis referred so as replica-symmetric (RS).\\
In~\cite{15RUYZ} it is discussed how this is not true in all cases, i.e. there exist a regime wherein the glass basin undergoes an ergodicity breaking and fractures into sub-basins. Nevertheless, here we stick to the simple RS ansatz. In this case, since the action $S(Q_{ab})$ must in turn be invariant for any replica permutations, the most general form that the Hessian $M$ can take is
\begin{eqnarray}
 M_{ab;cd} &=& M_1 \big(\frac{\delta_{ac}\delta_{bd} + \delta_{ad}\delta_{bc}}{2}\big) \nonumber\\&&+M_2 \big(\frac{\delta_{13} +\delta_{bd} + \delta_{ad} +\delta_{bc}}{4}\big) + M_3,
\end{eqnarray}
and the same goes for the covariance matrix $G_{ab;cd}$. This form is completely general as it only pertains to the RS symmetry; then the only model-dependence is in the parameters $M_1$, $M_2$ and $M_3$, which must be computed case by case and are generally dependent on the external parameters like temperature or magnetic field.

The diagonalization of the tensor $M_{ab;cd}$ is an exercise of standard linear algebra and has been already carried out many times, see for example~\cite{CS92,BM79,98DKT} and~\cite{Z10} where it is proposed as an exercise. It is found that the tensor $M$ has only three distinct eigenvalues
\begin{eqnarray}
 \lambda_R &=& M_1\\
 \lambda_L &=& M_1 + (s-1)(M_2+sM_3)\\
 \lambda_A &=& M_1 + \frac{s-2}{2}M_2,
\end{eqnarray}
and the same goes for the tensor $G$. Those three eigenvalues (or modes) are called the \emph{replicon}, \emph{longitudinal}, and \emph{anomalous}, respectively~\cite{Z10}.
We are interested in the longitudinal mode (which in the limit $s\to 0$ is degenerate with the anomalous one), which becomes soft at the yielding transition~\cite{16RU,16UZ}. Let us consider the $G$ tensor. Because of replica symmetry, there are only three distinct correlators that one can define, namely
\begin{eqnarray}
 G_{12;12} &=& \frac{G_1}{2} + \frac{G_2}{2} + G_3\\
 G_{12;13} &=& \frac{G_2}{4} + G_3\\
 G_{12;34} &=& G_3
\end{eqnarray}
and in the limit $s\to0$ we know that
\begin{equation}
 \frac{1}{\lambda_L} = G_1-G_2.
\end{equation}
It is then immediate to check that
\begin{eqnarray}
 G_{12;12} -2G_{12;13} + G_{12;34} &=& \frac{G_1}{2} \propto \frac{1}{\lambda_R} \equiv G_R\\
 G_{12;12} -4G_{12;13} + 3G_{12;34} &=& \frac{G_1-G_2}{2} \propto \frac{1}{\lambda_L} \equiv G_L
\end{eqnarray}
which then implies
\begin{equation}
 G_L(\boldsymbol{r}) = 2G_R(\boldsymbol{r}) -\Gamma_2(\boldsymbol{r}),
\end{equation}
with the definitions
\begin{eqnarray}
 G_R(\boldsymbol{r}) &\equiv& \overline{{\left<Q_{ab}(r)Q_{ab}(0)\right>}} - 2\overline{{\left<Q_{ab}(r)Q_{ac}(0)\right>}} \nonumber\\&+& \overline{{\left<Q_{ab}(r)\right>\left<Q_{cd}(0)\right>}} \label{eq:defGR}\\
 \Gamma_2(\boldsymbol{r}) &\equiv& \overline{{\left<Q_{ab}(\boldsymbol{r})Q_{ab}(0)\right>}} -\overline{{\left<Q_{ab}(\boldsymbol{r})\right>\left<Q_{ab}(0)\right>}}, \label{eq:defG2}
\end{eqnarray}
as in the main text. We have used $\Gamma_2 = G_{12;12} - G_{12;34}$ which derives from replica symmetry, as $\left<Q_{12}\right>\left<Q_{12}\right> = \left<Q_{12}Q_{34}\right>$ in the replica-symmetric phase.

Let us now detail how to transform these definitions into quantities that can be measured in simulation. We start by "localizing" the definition of the $Q_{ab}$ overlap in the following way
 \begin{equation}
  Q_{ab}(\B r) \equiv \sum_{i=1}^N\theta(\ell-|\B r_i^a-\B r_i^b|)\delta (\B r -\B r_i^a).
  \label{eq:defQr}
 \end{equation}
 In a thermal simulation the $a$ and $b$ configurations would depend on the time $t$, and so would the $Q_{ab}(r)$, so one would need to perform the in-state thermal average $\left<\bullet\right>$ by considering the equilibrium value of these quantities. In the present paper we focus un athermal solids under quasi-static shear, so we do not have dynamics and the $a$ and $b$ configurations will simply be two distinct minima of the inter-particle potential obtained through the protocol described in the main text, and the thermal average will be the average over this ensemble of configurations which make up a glassy patch.\\
We now apply the definition \eqref{eq:defQr} in the \eqref{eq:defGR}, \eqref{eq:defG2} to construct the correlators. For illustrative purposes we use the $\Gamma_2(\boldsymbol{r})$. We get, omitting the overline to lighten the notation,
\begin{widetext}
\begin{equation}
 \left<(Q_{ab}(\boldsymbol{x})-\left<Q_{ab}(\boldsymbol{x})\right>)(Q_{ab}(\boldsymbol{x}+\boldsymbol{r})-\left<Q_{ab}(\boldsymbol{x}+\boldsymbol{r})\right>)\right> = \sum_{ij}[(u^{ab}_i  -Q_{ab}) (u^{ab}_j - Q_{ab})]\delta (\B r + \boldsymbol{x} -\B r_i^a)\delta (\B x -\B r_j^a),
\end{equation}
\end{widetext}
with
 \begin{equation}
 u^{ab}_i \equiv \theta(\ell-|\boldsymbol{r}_i^a-\boldsymbol{r}_i^b|),
 \end{equation}
 as in the main text, and we used that $\left<Q_{ab}(x)\right> = Q_{ab}$. Because of translational invariance, the correlator is actually independent of $\boldsymbol{x}$. We can get rid of $\boldsymbol{x}$ by performing an integration over this variable, which, using the $\delta$-functions, gives as a result
\begin{equation}
\sum_{ij}[(u^{ab}_i  -Q_{ab}) (u^{ab}_j - Q_{ab})]\delta (\B r - (\B r_i^a- \boldsymbol{r}_j^a))
\end{equation}
then, following~\cite{16BCJPSZ}, we omit the terms with $i=j$ (which are anyway relevant only for $\boldsymbol{r} = 0$) and we normalize the correlator with the pair distribution function of the glass; we finally obtain
\begin{equation}
\frac{ \sum_{i\neq j}(u^{ab}_i-Q_{ab}) (u^{ab}_j-Q_{ab})\delta(\boldsymbol{r}-(\boldsymbol{r}_{i}^a-\boldsymbol{r}_{j}^a)) }{ \sum_{i\neq j}\delta(\boldsymbol{r}-(\boldsymbol{r}_{i}^a-\boldsymbol{r}_{j}^a)) } \equiv \tilde \Gamma_2(\B r),
\end{equation}
as in the main text. The derivation for the $\tilde G_R(\boldsymbol{x})$ is then an obvious generalization.

%


\end{document}